\documentclass[pra,twocolumn,amsmath,amssymb,superscriptaddress]{revtex4-1}
\usepackage{epsfig,amsmath}
\usepackage{subfigure}
\usepackage{graphicx}
\usepackage{dcolumn}
\usepackage{stmaryrd}
\usepackage{mathrsfs}
\usepackage{pifont}
\usepackage{amsthm}
\usepackage{amssymb}
\usepackage{bm}
\usepackage{latexsym}
\usepackage[colorlinks=true,linkcolor=blue,citecolor=blue]{hyperref}
\usepackage{color}
\usepackage{epstopdf}

\begin{document}

\title{Floquet generation of magnonic NOON state}

\author{Shi-fan Qi}
\affiliation{School of Physics, Zhejiang University, Hangzhou 310027, Zhejiang, China}

\author{Jun Jing}
\email{jingjun@zju.edu.cn}
\affiliation{School of Physics, Zhejiang University, Hangzhou 310027, Zhejiang, China}

\date{\today}

\begin{abstract}
We propose a concise and deterministic protocol to generate NOON states in a hybrid system consisting of a superconducting qubit, a circuit resonator mode, and two magnonic modes, based on Floquet engineering. In particular, we construct a time-reversal-symmetry broken Hamiltonian for chiral state propagation of the three continuous-variable modes depending on qubit state, by the time modulation over qubit-resonator interaction and magnon frequency. Then an arbitrary magnonic NOON state can be generated by a typical preparing-and-measurement procedure. We analyze the robustness of our protocol against the systematic errors in the qubit-magnon coupling strength, the Floquet-driving intensity, the frequency mismatch of the magnons, and the counter-rotating interactions. We can obtain a high-fidelity NOON state in the presence of the quantum dissipation on all components.
\end{abstract}

\maketitle

\section{Introduction}

NOON states, i.e., $(|N0\rangle+|0N\rangle)/\sqrt{2}$ with $N$ integer, consisting of two symmetric components in the maximal superposition~\cite{quaninter} constitute a prominent class of highly entangled states~\cite{qe}. They offer diversified applications in quantum metrology~\cite{quantmetro}, quantum communication~\cite{qco}, and quantum information processing~\cite{quantinf}. The NOON-state generation protocols~\cite{noon1,noon2,noon3,noon4,noon5,noon6,noon7,noon8,noon9,triplenoon} are conventionally developed on various Rabi oscillations. They have been realized in multiple quantum platforms, such as polarization states of photons~\cite{polarnoon}, nuclear spin of molecules~\cite{magnetsense}, optical paths of photons~\cite{lightnoon}, ultracold dipolar atoms in an optical superlattice setup~\cite{atominter,designnoon}, excitations in superconducting resonators~\cite{noon1,noon3}, and phonons in ion trap~\cite{ionnoon}. The ultra-precise control over complex quantum devices and decoherence of quantum systems~\cite{noon10}, however, make it extremely difficult to create a NOON state with a large $N$. It remains therefore interesting to find fast and faithful approaches to generate NOON states in low-decoherence systems. Here we propose to realize a magnonic NOON state by virtue of the chiral state transfer based on Floquet engineering.

The magnonic system is a growing field of research on magnetic devices that operate in the quantum realm. With unique properties such as high tunability, long coherent-time, and strong dipole-dipole coupling to the microwave photons and qubits, the magnonic modes have been used as the information carrier in an even broader variety of hybrid systems~\cite{magnon,magnon2,magnon3}. They are thus capable to prepare and manipulate various nonclassical states~\cite{nanocavity,size,mppentangle,bistability,cavitymagnonmechan,magnonqubit1,magnonqubit2,yigcavity1,yigcavity2}. Bell states of the magnon-photon system can be observed in both theory proposal~\cite{yigcavitybell2} and experimental demonstration~\cite{yigcavitybell}. Analogous to the cavity quantum electrodynamics~\cite{circuit}, Ref.~\cite{magnonqubit} proposed a magnonic cat-state generation protocol, in which the magnon was directly and quantum-coherently coupled to a superconducting transmon qubit.

In a broader view, NOON states could be categorized to the nonclassical states in (a non-normalized) form of $|\varphi\rangle|0\rangle+|0\rangle|\varphi\rangle$, where $|\varphi\rangle$ is an arbitrary pure state. They can result from a chiral state transfer depending on the symmetrically superposed state of a two-state system in charge of control. Chirality~\cite{chiral,chiral2} has been found to play an important role in the fractional quantum Hall effect~\cite{fraction} in the magnetic materials. It breaks the time reversal symmetry~\cite{timereversal} and perfectly realizes directional state transfer~\cite{chiralcurrent,floquetnoon}. In recent experiments~\cite{chiralspin,chiralspin2} on superconducting circuits and qubits, a three-spin interaction with chirality is fabricated by specially designed Floquet driving. Floquet engineering by fast periodic modulation over the characteristic frequency of a quantum system is a major control approach to the desired effective Hamiltonian for the long-time dynamics of the system~\cite{Floquet1,James,Floquettheory,Floquet2}. Also it has been implemented to realize quantum switch~\cite{switch}, chiral ground state current~\cite{chiralcurrent}, and quantum simulation~\cite{Floquetsimu}.

In this work, we consider a hybrid qubit-resonator-magnon system upon a state-of-the-art device, in which the qubit is coupled to the resonator mode with time-modulated strength~\cite{timecoupling,chiralcurrent} and simultaneously coupled to two magnonic modes~\cite{magnonqubit}. Floquet engineering is applied to the frequencies of the magnonic modes~\cite{floquetmagnon}. Using appropriate frequencies, intensities, and local phases in control, we can fabricate an effective time-reversal-symmetry broken Hamiltonian to ensure chiral state transfer amongst resonator and magnonic modes. The transfer direction depends on the state of the qubit. A magnonic NOON state can thus be generated upon preparing the resonator mode as a Fock state $|N\rangle$~\cite{Fockmeasure}.

The rest part of this work is structured as follows. In Sec.~\ref{secmodel}, we introduce the full Hamiltonian for a hybrid qubit-resonator-magnon system and then derive the effective Hamiltonian for a perfect chiral state-transfer amongst the three continuous-variable components. In Sec.~\ref{syserror}, we discuss the systematic errors from the qubit-magnon coupling strength, the Floquet driving intensity, the frequency mismatch of the two magnonic modes, and the counter-rotating interactions. The NOON-state generation protocol and its fidelity are analyzed in Secs.~\ref{prepar} and \ref{fideana}, respectively. The whole work is summarized in Sec.~\ref{conclu}.

\section{Model and chiral state transfer}\label{secmodel}

\begin{figure}[htbp]
\centering
\includegraphics[width=0.45\textwidth]{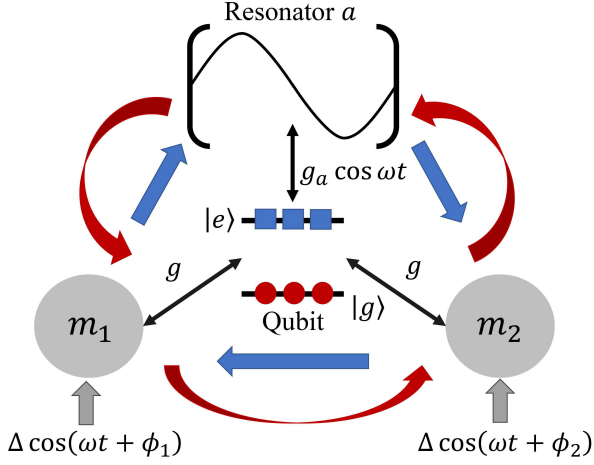}
\caption{Diagram of the chiral state propagations amongst three continuous-variable modes $a$, $m_1$, and $m_2$ in a hybrid qubit-resonator-magnon system. The qubit population on the excited and ground states determines respectively the clockwise and anticlockwise directions of the chiral state propagation.}\label{model}
\end{figure}

Consider a hybrid system that consists of a circuit resonator, a superconducting qubit, and two YIG spheres in their ground states (the Kittel mode of the spin ensemble), as shown in Fig.~\ref{model}. The transition frequencies of the qubit, resonator, and magnon mode-$k$ are supposed to be resonant with each other in the microwave regime, i.e., $\omega_q=\omega_a=\omega_k$. The qubit-resonator interaction and the frequencies of the two magnonic modes are under the Floquet engineering~\cite{timecoupling,chiralcurrent,floquetmagnon}. Then in the rotating frame with respect to the free Hamiltonian $H=\omega_a(\sigma^+\sigma^-+a^\dag a+\sum^2_{k=1}m^\dag_km_k)$, the full Hamiltonian can be written as
\begin{equation}\label{Hamiltonian}
\begin{aligned}
H&=g_a\cos(\omega t)\left(\sigma^+a+\sigma^-a^\dag\right) \\
+&\Delta\sum_{k=1}^2\cos(\omega t+\phi_k)m^\dag_km_k+g\sum_{k=1}^2\left(\sigma^+m_k+\sigma^-m^\dag_k\right),
\end{aligned}
\end{equation}
where $a$ ($a^{\dagger}$) and $m_k$ ($m_k^{\dagger}$) are the annihilation (creation) operators for the resonator mode and the $k$th magnonic mode, respectively, and $\sigma^+\equiv|e\rangle\langle g|$ and $\sigma^-\equiv|g\rangle\langle e|$ are Pauli transition operators. The qubit-resonator interaction is characterized by the coupling strength $g_a$ and the driving frequency $\omega$~\cite{timecoupling,chiralcurrent}. For magnons, $\Delta$, $\omega$, and $\phi_k$ represent the Floquet driving intensity, frequency, and local phases, respectively~\cite{floquetmagnon}. $g$ is the qubit-magnon coupling strength~\cite{magnonqubit}. The resonator-magnon interaction is vanishing when they depart with a significant distance.

In the rotating frame with respect to the magnon Hamiltonian under driving [the second term in Eq.~(\ref{Hamiltonian})], we have
\begin{equation}\label{Ham1}
\begin{aligned}
H(t)&=g_a\cos(\omega t)\left(\sigma^+a+\sigma^-a^\dag\right) \\
&+g\sum_{k=1}^2\left(e^{if[\sin\phi_k-\sin(\omega t+\phi_k)]}\sigma^+m_k+{\rm H.c.}\right),
\end{aligned}
\end{equation}
where $f\equiv\Delta/\omega$ is the ratio of the driving intensity and frequency. According to the perturbative expansion ordered by the Bessel functions of the first kind, i.e., $e^{iz\sin y}=\sum_{n=-\infty}^{n=+\infty}J_n(z)e^{iny}$, we have
\begin{equation}\label{HI}
\begin{aligned}
H_I&=H_a\left(e^{i\omega t}+e^{-i\omega t}\right)+H_0\\
&+\sum^2_{k=1}\sum^{\infty}_{n=1}\left[H_n^{(k)}e^{in\omega t}+H_{-n}^{(k)}e^{-in\omega t}\right],
\end{aligned}
\end{equation}
where
\begin{equation}\label{Hak}
\begin{aligned}
H_a&=\frac{g_a}{2}\left(\sigma^+a+\sigma^-a^\dag\right),\\
H_0&=gJ_0(f)\sum_{k=1}^2\left(\sigma^+m_k+\sigma^-m^\dag_k\right),\\
H_{n}^{(k)}&=ge^{-if\sin\phi_k}e^{in\phi_k}J_n(f)\sigma^- m^\dag_k\\
&+(-1)^nge^{if\sin\phi_k}e^{in\phi_k}J_n(f)\sigma^+m_k,
\end{aligned}
\end{equation}
And $H_{-n}^{(k)}$ is the Hermitian conjugate of $H_n^{(k)}$. Using the James' method~\cite{James,Floquettheory}, the interaction Hamiltonian $H_I$ can therefore be written as
\begin{equation}\label{effHam0}
\begin{aligned}
H_I&\approx H_0+\frac{1}{\omega}\sum_{k=1}^2\left\{\left[H_a, H^{(k)}_{-1}\right]+\left[H_{1}^{(k)}, H_a\right]\right\}\\
&+\sum_{k=1}^2\frac{1}{n\omega}\sum^{\infty}_{n=1}\left[H^{(k)}_n, H^{(3-k)}_{-n}\right],
\end{aligned}
\end{equation}
up to the order of $\mathcal{O}(1/\omega^2)$. The zeroth-order term $H_0$ describes the effective qubit-magnon interaction under Floquet driving, whose coupling-strength $gJ_0(f)$ can be tuned by varying the ratio $f$. This term is phase-independent and can be eliminated by setting $J_0(f)=0$, i.e., $f\approx2.4048$. In this situation, we have a $\sigma_z$-dependent effective Hamiltonian
\begin{equation}\label{Heff}
\begin{aligned}
H_{\rm eff}&=\sigma_z\Big[a^\dagger\left(g_1e^{if\sin\phi_1}m_1+g_2e^{if\sin\phi_2}m_2\right)\\
&-ig_{12}e^{if(\sin\phi_2-\sin\phi_1)}m^\dag_1m_2+{\rm H.c.}\Big],
\end{aligned}
\end{equation}
where the coupling strengths are
\begin{equation}\label{cpstr}
\begin{aligned}
g_k&=-\frac{g_ag}{\omega}J_1(f)\cos\phi_k, \quad k=1,2, \\
g_{12}&=\frac{2g^2}{\omega}\sum^{\infty}_{n=1}\frac{J^2_n(f)}{n}\sin\left[n(\phi_2-\phi_1)\right].
\end{aligned}
\end{equation}

To render a perfect chiral transfer amongst the three components $a$, $m_1$ and $m_2$, it is required that $|g_1|=|g_2|=|g_{12}|=g_{\rm eff}$, i.e., $|\cos(\phi_1)|=|\cos(\phi_2)|$. Note this condition implies the distinction between driving the coupling strength and driving the frequency in realizing the three-body chirality. In previous works~\cite{floquetnoon,chiralspin2} for generating a chiral state transfer by Floquet engineering on the three components' frequencies, the driving phases have to be uniformly distributed in $[0, 2\pi]$. While in our protocol (as a mixture of driving both frequency and coupling strength), the phases $\phi_k$ are not fixed. A non-unique solution is $\phi_1=2\pi/3$ and $\phi_2=4\pi/3$, and then
\begin{equation}\label{ga}
g_a=\frac{4g}{J_1(f)}\sum^{\infty}_{n=1}\frac{J^2_n(f)}{n}\sin\left(\frac{2n\pi}{3}\right).
\end{equation}
Consequently, the effective Hamiltonian in Eq.~(\ref{Heff}) could be written in a Matrix-product formation,
\begin{equation}\label{Heffmatrix}
\begin{aligned}
H_{\rm eff}&=\sigma_z\\
\times&\left[a^\dag, m^\dag_1, m^\dag_2\right]g_{\rm eff}\begin{bmatrix}
0 & e^{\frac{i\sqrt{3}f}{2}} & e^{-\frac{i\sqrt{3}f}{2}}\\
e^{-\frac{i\sqrt{3}f}{2}} & 0 &-ie^{-i\sqrt{3}f}\\
e^{\frac{i\sqrt{3}f}{2}} &ie^{i\sqrt{3}f} & 0
\end{bmatrix}\begin{bmatrix}
a\\m_1\\m_2\end{bmatrix},
\end{aligned}
\end{equation}
which is convenient to obtain the time evolution of these continuous-variable operators
\begin{equation}\label{operatoramm}
\begin{bmatrix}
a(t) \\ m_1(t) \\ m_2(t)
\end{bmatrix}=\hat{T}(t)\begin{bmatrix}
a(0) \\ m_1(0) \\ m_2(0)
\end{bmatrix}.
\end{equation}
Here $\hat{T}(t)$ depends on the qubit state. When the qubit is in the excited state $|e\rangle$, we have
\begin{equation}\label{operatorTe}
\begin{aligned}
&\hat{T}(t)=\hat{T}^{(e)}(t)\\
&=\frac{1}{3}\begin{bmatrix}
x(t) & -ie^{\frac{i\sqrt{3}f}{2}}y(t) & ie^{-\frac{i\sqrt{3}f}{2}}z(t)\\
ie^{-\frac{i\sqrt{3}f}{2}}z(t) & x(t)& -e^{-i\sqrt{3}f}y(t)\\
-ie^{\frac{i\sqrt{3}f}{2}}y(t) & -e^{i\sqrt{3}f}z(t) & x(t)
\end{bmatrix},
\end{aligned}
\end{equation}
where
\begin{equation}\label{xyz}
\begin{aligned}
x(t)&=1+2\cos\left(\sqrt{3}g_{\rm eff}t\right),\\
y(t)&=1+2\cos\left(\sqrt{3}g_{\rm eff}t-\frac{2\pi}{3}\right),\\
z(t)&=1+2\cos\left(\sqrt{3}g_{\rm eff}t+\frac{2\pi}{3}\right).
\end{aligned}
\end{equation}
The transformation matrix $\hat{T}^{(e)}$ entails a sufficient condition for a clockwise and periodic chirality. It is interesting to find that $m_2(t)=a(0)$, $m_1(t)=m_2(0)$, and $a(t)=m_1(0)$ when $t=(2\pi/3+2n\pi)/(\sqrt{3}g_{\rm eff})$; $m_2(t)=m_1(0)$, $m_1(t)=a(0)$, and $a(t)=m_2(0)$ when $t=(4\pi/3+2n\pi)/(\sqrt{3}g_{\rm eff})$; and $m_k(t)=m_k(0)$, $a(t)=a(0)$ when $t=2n\pi/(\sqrt{3}g_{\rm eff})$ with $n$ integer. This transfer is exactly the rotation $a\to m_2\to m_1\to a$ described by the straight-line arrows in Fig.~\ref{model}, corresponding to a clockwise chiral propagation of states in the Schr\"odinger picture, i.e., $|\varphi_a\varphi_1\varphi_2\rangle \to |\varphi_1\varphi_2\varphi_a\rangle \to |\varphi_2\varphi_1\varphi_a\rangle$, where $|\varphi_a\rangle$ and $|\varphi_k\rangle$ are arbitrary states for the resonator and magnonic mode-$k$, respectively.

Suppose that the resonator is prepared as an arbitrary superposed state $C_n|n\rangle$ and the two magnonic modes are in their ground states, i.e.,
\begin{equation}\label{varphi0}
|\varphi(0)\rangle=\sum_nC_n|n00\rangle=\sum_n\frac{C_n}{\sqrt{n!}}\left(a^{\dag}\right)^n|000\rangle.
\end{equation}
By virtue of Eq.~(\ref{operatorTe}), it is straightforward to express the time-evolved state as
\begin{equation}
\begin{aligned}
|\varphi(t)\rangle&=\sum_n\frac{C_n}{\sqrt{n!}}\Big[a^\dag\hat{T}^{(e)\dagger}_{11}(t)
+m_1^\dag\hat{T}^{(e)\dagger}_{12}(t)\\ &+m_2^\dag\hat{T}^{(e)\dagger}_{13}(t)\Big]^n|000\rangle.
\end{aligned}
\end{equation}
For example, when $|\varphi(0)\rangle=|100\rangle$, we have
\begin{equation}
\begin{aligned}
|\varphi(t)\rangle&=\frac{1}{3}\Big[x(t)|100\rangle-ie^{\frac{i\sqrt{3}f}{2}}z(t)|010\rangle\\
&+ie^{\frac{-i\sqrt{3}f}{2}}y(t)|001\rangle\Big].
\end{aligned}
\end{equation}

\begin{figure}[htbp]
\centering
\includegraphics[width=0.45\textwidth]{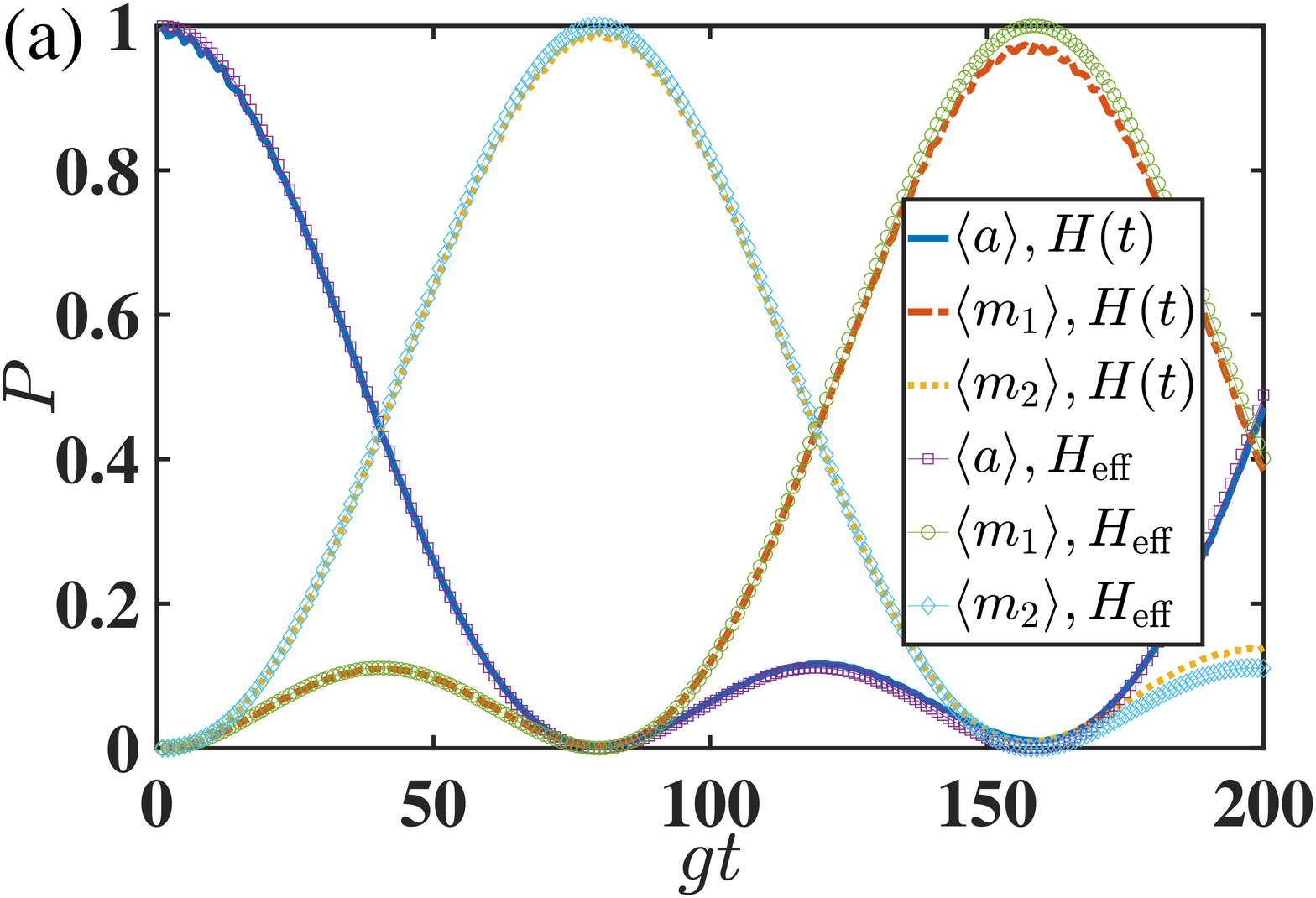}
\includegraphics[width=0.45\textwidth]{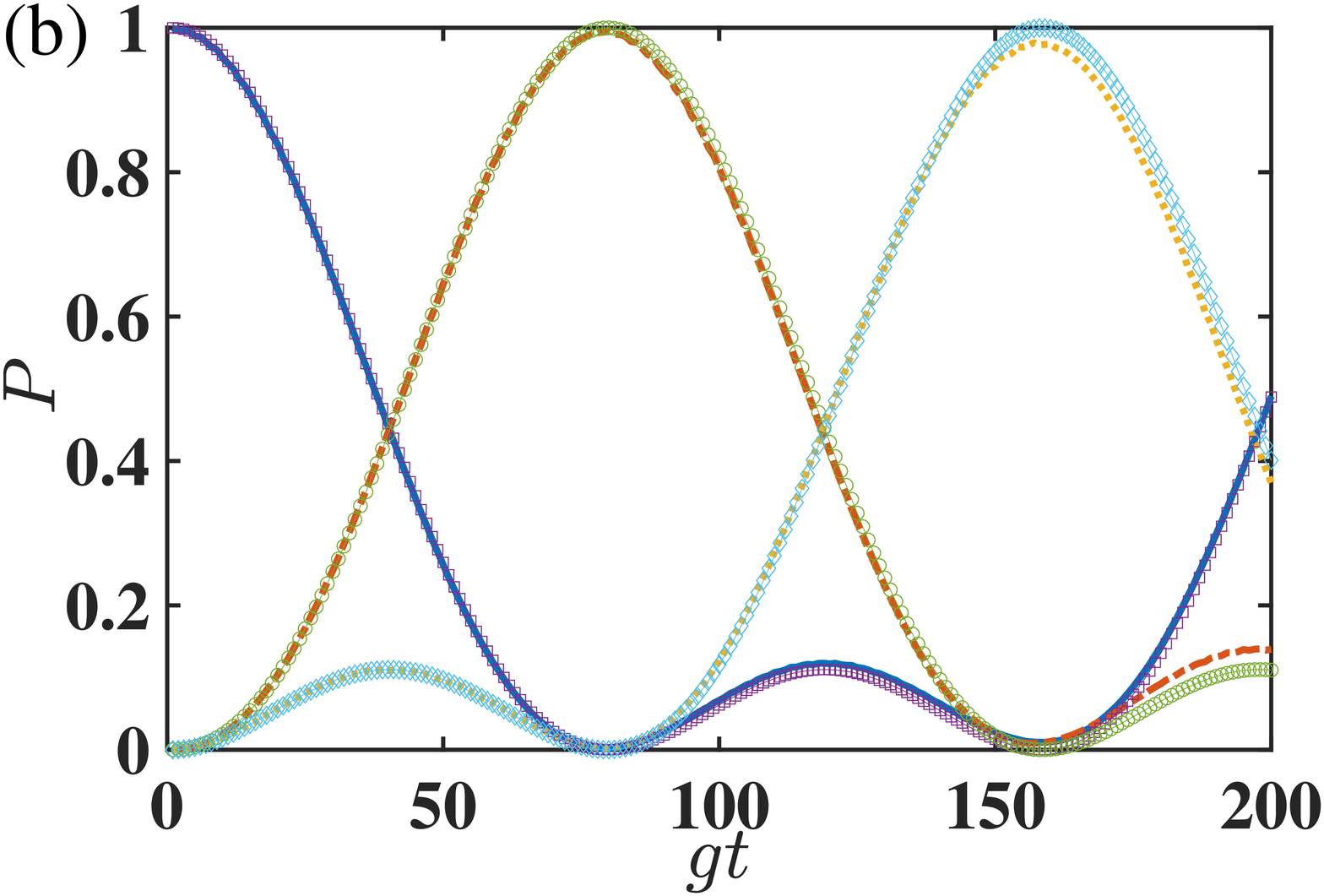}
\caption{Chiral dynamics of the state populations $P_j$ for the three modes, $j=a,1,2$, by numerical simulation with the Floquet engineering Hamiltonian~(\ref{Hamiltonian}) and analytical evaluation with the effective Hamiltonian~(\ref{Heff}). In (a), the qubit is at $|e\rangle$ and in (b), it is at $|g\rangle$. The three modes are initialized as $|1\rangle_a|0\rangle_{m_1}|0\rangle_{m_2}$. The parameters are set as $\omega=20g$, $f=\Delta/\omega=2.4048$, $\phi_1=2\pi/3$, and $\phi_2=4\pi/3$.}\label{result}
\end{figure}

We can have a unit transfer fidelity at the desired moments for the general superposed state in Eq.~(\ref{varphi0}), e.g., when $t=2\pi/(3\sqrt{3}g_{\rm eff})$, $|\varphi(t)\rangle=\sum_nC_n|00n\rangle$. To avoid the influence from local phases~\cite{localphase} during the whole evolution, the state-transfer fidelity can be measured by time-dependent state populations
\begin{equation}\label{popu}
P_j(t)=\sum_{C_n\neq0}|\langle\varphi(t)|n\rangle_j|^2,
\end{equation}
where $|n\rangle_j$, $j=a,1,2$, indicates that the marked mode is in the Fock state $|n\rangle$ and the other two modes are in their ground states.

In parallel to Eq.~(\ref{operatorTe}),we have
\begin{equation}\label{operatorTg}
\begin{aligned}
&\hat{T}(t)=\hat{T}^{(g)}(t)\\
&=\frac{1}{3}\begin{bmatrix}
x(t) & -ie^{\frac{i\sqrt{3}f}{2}}z(t) & ie^{-\frac{i\sqrt{3}f}{2}}y(t)\\
ie^{-\frac{i\sqrt{3}f}{2}}y(t) & x(t)& -e^{-i\sqrt{3}f}z(t)\\
-ie^{\frac{i\sqrt{3}f}{2}}z(t) & -e^{i\sqrt{3}f}y(t) & x(t)
\end{bmatrix},
\end{aligned}
\end{equation}
when the qubit is in the ground state $|g\rangle$. The transformation matrix $\hat{T}^{(g)}$ indicates an anticlockwise chirality, yielding the transfer along the rotation $a\to m_1\to m_2\to a$ described by the curved arrows in Fig.~\ref{model}. Then in the Schr\"odinger picture, a chiral evolution emerges as $|\varphi_a\varphi_1\varphi_2\rangle \to |\varphi_2\varphi_a\varphi_1\rangle \to |\varphi_1\varphi_2\varphi_a\rangle$. For the same initial states in Eq.~(\ref{varphi0}), we have
\begin{equation}
\begin{aligned}
|\varphi(t)\rangle&=\sum_n\frac{C_n}{\sqrt{n!}}\Big[a^\dag\hat{T}^{(g)\dagger}_{11}(t)
+m_1^\dag\hat{T}^{(g)\dagger}_{12}(t)\\ &+m_2^\dag\hat{T}^{(g)\dagger}_{13}(t)\Big]^n|000\rangle.
\end{aligned}
\end{equation}
by virtue of Eq.~(\ref{operatorTg}).

Figures~\ref{result}(a) and \ref{result}(b) show the time-evolved state populations $P_j(t)$ under the effective Hamiltonian~(\ref{Heff}) (see the lines with markers) and the system Hamiltonian in Eq.~(\ref{Hamiltonian}) (see the lines with no markers), demonstrating respectively the clockwise and anticlockwise chirality. The numerical and analytical results are found to match perfectly with each other. The qubit state determines the chiral direction. A superposed state of the qubit gives rise to two chiral directions, which are demanded to generate the NOON state. The period of the chiral state propagation is state-independent and uniquely determined by the coupling strengths in Eq.~(\ref{cpstr}) under the condition $J_0(f)=0$.

\section{Systematic errors}\label{syserror}

In practice, the ideal chiral state transfer in Fig.~\ref{result} using the effective Hamiltonian~(\ref{Heff}) cannot be exactly realized because of the imperfections and constraints in experiments. Even under the desired conditions in the last section, i.e., fast engineering, exact cancellation of the phase-independent term, and accurate setting of the coupling strength, it is inevitable to estimate the perturbative effect on the transport protocols from the fluctuations in parameters of the full Hamiltonian. In the rest of this section, we analyze the systematic errors induced by the qubit-magnon coupling-strength deviation, the unstable Floquet driving, the mismatch of the magnon frequencies, and the presence of the counter-rotating interactions.

\subsection{The qubit-magnon coupling-strength deviation}

\begin{figure}[htbp]
\centering
\includegraphics[width=0.45\textwidth]{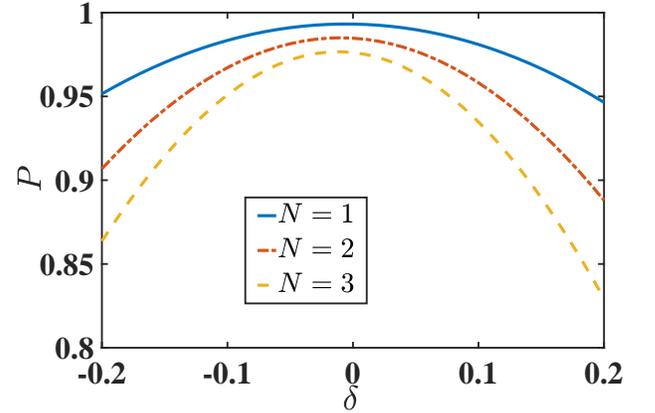}
\caption{Chiral state-transfer fidelity of the target state $(|e00N\rangle+|g0N0\rangle)/\sqrt{2}$ under the nonideal Hamiltonian~(\ref{Hamiltoniansys}) as a function of the systematic error associated with the qubit-magnon coupling-strength. Ideally (under the effective Hamiltonian) the initial state $(|eN00\rangle+|gN00\rangle)/\sqrt{2}$ would become the target state at the desired moment $T=2\pi/(3\sqrt{3}g_{\rm eff})$. Here the parameters are set the same as Fig.~\ref{result}.}\label{errordelta}
\end{figure}

The coupling strengths between qubit and magnonic modes [the last term in Eq.~(\ref{Hamiltonian})] are chosen the same in magnitude. They depend practically on the distance between the YIG spheres and the qubit. We first consider their deviation from a fixed value. The system Hamiltonian can thus be rewritten as
\begin{equation}\label{Hamiltoniansys}
\begin{aligned}
H'&=g_a\cos(\omega t)(\sigma^+a+\sigma^-a^\dag)\\
&+\Delta\cos(\omega t+\phi_1)m^\dag_1m_1+g(1+\delta)(\sigma^+m_1+\sigma^-m^\dag_1)\\
&+\Delta\cos(\omega t+\phi_2)m^\dag_2m_2+g(1-\delta)(\sigma^+m_2+\sigma^-m^\dag_2),
\end{aligned}
\end{equation}
where $\delta$ represents the magnitude of the relative error. Under the same settings that $\phi_1=2\pi/3$, $\phi_2=4\pi/3$, $J_0(f=\Delta/\omega)=0$, and $|g_k|=|g_{12}|$ in Eq.~(\ref{cpstr}) for the chiral transfer, the effective Hamiltonian in Eq.~(\ref{Heffmatrix}) is modified by changing the coefficient matrix into
\begin{equation}\label{Heffsys}
g_{\rm eff}\begin{bmatrix}
0 & (1+\delta)e^{\frac{i\sqrt{3}f}{2}} & (1-\delta)e^{-\frac{i\sqrt{3}f}{2}}\\
(1+\delta)e^{-\frac{i\sqrt{3}f}{2}} & 0 &-i(1-\delta^2)e^{-i\sqrt{3}f}\\
(1-\delta)e^{\frac{i\sqrt{3}f}{2}} &i(1-\delta^2)e^{i\sqrt{3}f} & 0
\end{bmatrix}.
\end{equation}
It can be perturbatively decomposed into
\begin{equation}\label{Heffsyserror}
H'_{\rm eff}=H_{\rm eff}+\delta H_1+\delta^2H_2,
\end{equation}
where $H_{\rm eff}$ is the unperturbed effective Hamiltonian in Eq.~(\ref{Heffmatrix}). $H_1$ is the leading-order perturbation, whose coefficient matrix reads,
\begin{equation}\label{Heffsys1}
g_{\rm eff}\begin{bmatrix}
0 & e^{\frac{i\sqrt{3}f}{2}} & -e^{-\frac{i\sqrt{3}f}{2}}\\
e^{-\frac{i\sqrt{3}f}{2}} & 0 & 0\\
-e^{\frac{i\sqrt{3}f}{2}} & 0 & 0
\end{bmatrix}.
\end{equation}
Using the method in Ref~\cite{syserrorpo} and up to the second order of the systematic error $\delta$, one can obtain the population $p$ of the magnon-mode $m_2$ for the transfer $|e100\rangle\rightarrow|e001\rangle$ at the desired time $T=2\pi/(3\sqrt{3}g_{\rm eff})$ as
\begin{equation}\label{Pm}
p=1-\sum_{j=1}^2\left|\int^T_0dt\langle0|m_j(t)|\delta H_1|a^\dag(t)|0\rangle\right|^2=1-\delta^2,
\end{equation}
where $a(t)$ and $m_j(t)$ are the time-evolved operators given in Eq.~(\ref{operatoramm}) and $|0\rangle\equiv|0\rangle_a|0\rangle_{m_1}|0\rangle_{m_2}$ is the vacuum state. By the population definition in Eq.~(\ref{popu}), the state-transfer population $P$ for $|e00N\rangle$ from $|eN00\rangle$ can then be estimated as $P=p^N=(1-\delta^2)^N\approx 1-N\delta^2$. This result applies also to the state transfer $|gN00\rangle\rightarrow |g00N\rangle$. The nonideal dynamics under Hamiltonian~(\ref{Hamiltoniansys}) are provided in Fig.~\ref{errordelta} by the sensitivity of the transfer fidelity (population) of the target state $|\Phi(t)\rangle=(|e00N\rangle+|g0N0\rangle)/\sqrt{2}$ that evolves from the initial state $|\Phi(0)\rangle=(|eN00\rangle+|gN00\rangle)/\sqrt{2}$, to $\delta$. The populations are evaluated at $T$, that is determined by the ideal chiral state-transfers depending on Eqs.~(\ref{operatorTe}) and (\ref{operatorTg}). The state-transfer population in general declines with increasing $|\delta|$ and $N$. Our protocol is found to be robust even under $|\delta|\approx20\%$ and $N=3$, whereby the population is about $0.85$.

\begin{table}[htbp]
  \centering
  \begin{tabular}{|c|c|c|c|c|c|c|c|c|c|c|}
  \hline
  $N$ &1&2&3&4&5&6&7&8&9&10 \\
  \hline
  $P$ &0.98&0.97&0.95&0.93&0.90&0.87&0.84&0.80&0.76&0.72 \\
  \hline
\end{tabular}
\caption{Chiral state-transfer population of the target state $(|e00N\rangle+|g0N0\rangle)/\sqrt{2}$ under the total Hamiltonian in Eq.~(\ref{Hamiltoniansys}) with $\delta=0$. Here the parameters are set the same as Fig.~\ref{result}.}\label{NFidelity}
\end{table}

Note that $H_{\rm eff}$ is the second-order perturbation over $H$. So that one can find a less-than-unit population even when $\delta=0$. We list the results for various $N$ in Tab.~\ref{NFidelity}. The transfer population decreases monotonically with $N$ and it becomes less than $0.80$ when $N>8$. It is due to the fact that the second-order effective Hamiltonian is applicable in the regime $g\ll\omega$ and $g$ becomes practically $g\sqrt{N}$ for $|N\rangle$.

\subsection{The unstable Floquet driving}

To cancel the phase-independent term from the interaction Hamiltonian~(\ref{effHam0}) for a chiral propagation of quantum states, the ratio $f=\Delta/\omega$ of the Floquet-driving intensity $\Delta$ and the frequency $\omega$ is fixed to meet the requirement $J_0(f)=0$. We now estimate the effect of the control error arising from the Floquet-driving intensity, which is unstable in time domain. The error could then be regarded as random fluctuations. The Hamiltonian in Eq.~(\ref{Hamiltonian}) can thus be modified to
\begin{equation}\label{Hamiltonsto}
\begin{aligned}
H'&=g_a\cos(\omega t)\left(\sigma^+a+\sigma^-a^\dag\right)\\
&+\Delta\sum_{k=1}^2(1+\epsilon_k)\cos(\omega t+\phi_k)m^\dag_km_k\\
&+g\sum_{k=1}^2\left(\sigma^+m_k+\sigma^-m^\dag_k\right),
\end{aligned}
\end{equation}
where $\epsilon_k$ indicates a dimensionless factor for the driving intensity on the magnonic mode-$k$. It is assumed to be a random number in the range of $[0, \epsilon]$ with $\epsilon<1$ and $\epsilon_1\neq\epsilon_2$.

\begin{figure}[htbp]
\centering
\includegraphics[width=0.45\textwidth]{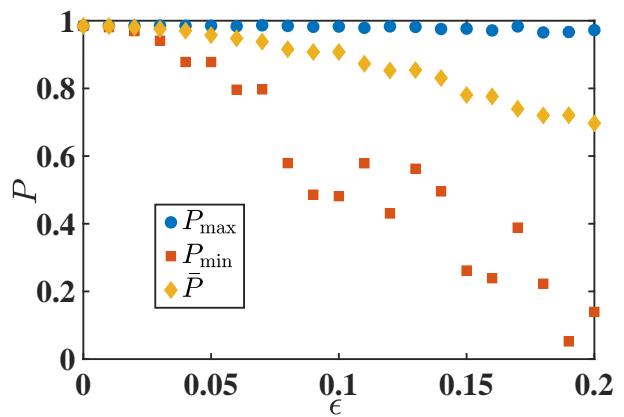}
\caption{Chiral state-transfer fidelity of the target state $(|e002\rangle+|g020\rangle)/\sqrt{2}$ under the nonideal Hamiltonian~(\ref{Hamiltonsto}) as a function of the systematic error associated with the Floquet-driving intensity. The circles, squares, and diamonds represent the maximum $P_{\rm max}$, minimum $P_{\rm min}$, and average populations $\bar{P}$, respectively. Here the parameters are set the same as Fig.~\ref{result}.}\label{Ffdelta}
\end{figure}

In Fig.~\ref{Ffdelta}, we present the sensitivity of the state-transfer population to the error upper-bound $\epsilon$. The initial state is chosen as $|\Phi(0)\rangle=(|e200\rangle+|g200\rangle)/\sqrt{2}$ and then the target state is $|\Phi[T=2\pi/(3\sqrt{3}g_{\rm eff})]\rangle=(|e002\rangle+|g020\rangle)/\sqrt{2}$. The populations, measured by the maximum $P_{\rm max}$, the minimum $P_{\rm min}$, and the average values $\bar{P}$, are obtained by $100$ numerical simulations using randomly distributed $\epsilon_k$'s. The distance between $P_{\rm max}$ and $P_{\rm min}$ is found to decrease roughly with increasing $\epsilon$. The average population $\bar{P}$ can be maintained above $0.90$ when $\epsilon\leq0.1$, and above $0.70$ when $\epsilon\leq0.2$. However, the minimum value $P_{\rm min}$ declines to below $0.50$ when $\epsilon$ approaches $0.1$. It implies the dramatic error caused by the fluctuation in the Floquet driving intensity, which is more significant than that in the qubit-magnon coupling-strength.

For a constant deviation $\epsilon_1=\epsilon_2=\epsilon$, we have an extra term in addition to the ideal effective Hamiltonian in Eq.~(\ref{Heff}), i.e.,
\begin{equation}\label{Heffepsilon}
H'_{\rm eff}=H_0+H_{\rm eff},
\end{equation}
where $H_0$ is the zeroth-order term in Eq.~(\ref{Hak}) and the ratio $f$ in both $H_0$ and $H_{\rm eff}$ is replaced with $f'=f(1+\epsilon)$. When $\epsilon=0.005$, the chiral state-transfer fidelity is found to be $P=0.84$ under $H'_{\rm eff}$ in Eq.~(\ref{Heffepsilon}) and it is about $P=0.82$ under the full Hamiltonian in Eq.~(\ref{Hamiltonsto}). When $\epsilon=0.01$, it drops to $P=0.50$ and $P=0.49$, respectively. Thus, the fidelity is sensitive to the error arisen from the unstable Floquet driving intensity. To render a perfect chiral state transfer among components $a$, $m_1$ and $m_2$, the magnitude of the zeroth-order should be kept as low as possible, i.e., $gJ_0(f')\ll g_{\rm eff}$.

\subsection{The frequency mismatch of magnons and qubit}

\begin{figure}[htbp]
\centering
\includegraphics[width=0.45\textwidth]{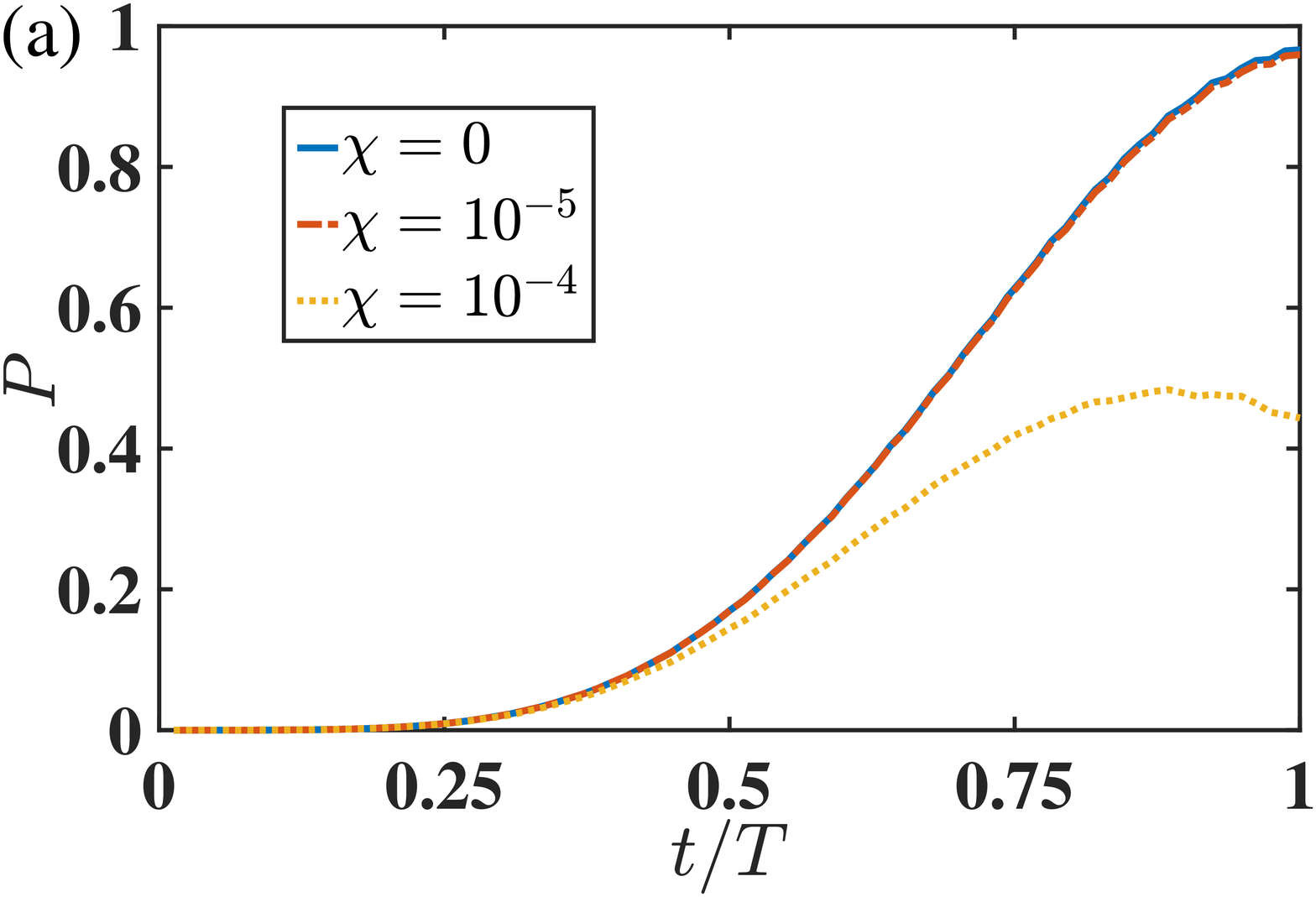}
\includegraphics[width=0.45\textwidth]{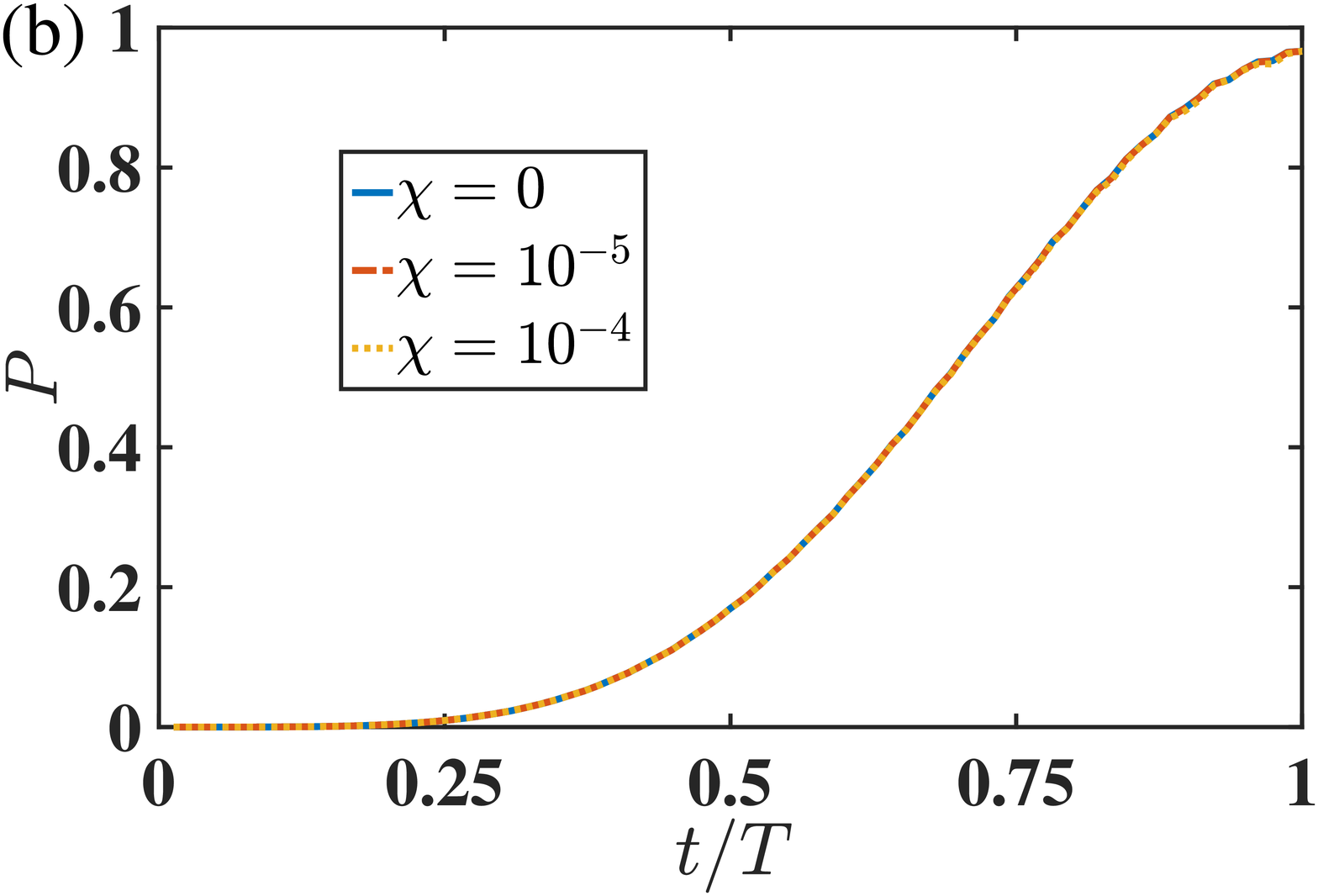}
\caption{(a) and (b): Chiral state-transfer population of the target state $(|e002\rangle+|g020\rangle)/\sqrt{2}$ under the deviated system Hamiltonians~(\ref{Htotsme}) and (\ref{Htotsqe}) with various error magnitude $\chi$, which are associated with the magnons and qubit frequency mismatch, respectively. Here the parameters are set the same as Fig.~\ref{result} and $T=2\pi/(3\sqrt{3}g_{\rm eff})$.}\label{mismatch}
\end{figure}

The four components in our system, i.e. the superconducting qubit, the resonator, and two magnonic modes are supposed to be resonant with each other in the microwave regime in Sec.~\ref{secmodel}. It is a challenging to achieve such accurate resonance in experiments. Therefore, the frequency mismatch effects on chiral state-transfer fidelity is important.

For the frequency mismatch of the two magnon modes, the deviated Hamiltonian in the Schr\"odinger picture could be written as
\begin{equation}\label{Htotsme}
H'=\omega_a(a^\dag a+\sigma^+\sigma^-)+\omega_a\sum^2_{k=1}(1+\chi_k)m^\dag_k m_k+H,
\end{equation}
where $\chi_1=\chi$, $\chi_2=-\chi$, $\chi$ represents the relative magnitude of the frequency mismatch, and $H$ is the Hamiltonian in Eq.~(\ref{Hamiltonian}). In Fig.~\ref{mismatch}(a), the chiral state transfer population dynamics from the initial state $|\Phi(0)\rangle=(|e200\rangle+|g200\rangle)/\sqrt{2}$ is presented under various $\chi$. One can observe that the population dynamics when $\chi=10^{-5}$ is close to the dynamics free of mismatch. And the target state population can be maintained above $0.95$. It is reduced to $0.4$ when $\chi=10^{-4}$. Nevertheless, our protocol is promising in the recent experiments~\cite{cavitymagnonics} since the bandwidths of the resonator and magnon modes are found to be around $10^{-6}\sim10^{-5}\omega_a$.

For the frequency deviation from the qubit, the Hamiltonian turns out to be
\begin{equation}\label{Htotsqe}
H'=\omega_aa^\dag a+\omega_a(1+\chi)\sigma^+\sigma^-+\omega_a\sum^2_{k=1}m^\dag_km_k+H.
\end{equation}
One can find in Fig.~\ref{mismatch}(b) that this deviation has no significant effect on the state transfer dynamics for the three lines under various $\chi$ are almost the same. It can be understood by the effective Hamiltonian, which is found to be $H'_{\rm eff}=\chi\omega_a\sigma^+\sigma^-+H_{\rm eff}$. The extra term about the qubit-frequency mismatch is clearly irrelevant to the chiral state transfer dynamics among the rest three bosonic modes.

\subsection{The presence of the counter-rotating interaction}

Back to the lab frame, the full Hamiltonian under the resonant condition without the rotating-wave approximation is written as
\begin{equation}\label{Hamcounter}
\begin{aligned}
H'&=\omega_a(a^\dag a+\sigma^+\sigma^-)\\ &+\sum^2_{k=1}\left[\omega_a+\Delta\cos(\omega t+\phi_k)\right]m^\dag_km_k\\
&+g_a\cos(\omega t)(\sigma^+a+\sigma^-a^\dag+\sigma^+a^\dag+\sigma^-a)\\
&+\sum_{k=1}^2g(\sigma^+m_k+\sigma^-m^\dag_k+\sigma^+m^\dag_k+\sigma^-m_k),
\end{aligned}
\end{equation}
where $\omega_a$ represents the characteristic frequency of the four components. In the Hamiltonian $H$~(\ref{Hamiltonian}) used for our chiral state-transfer protocol, both qubit-resonator and qubit-magnon coupling strengths $g_a$ and $g$ have to be much smaller than the transition frequency $\omega_a$ in magnitude. Although a faster speed of the chiral transfer favors a larger $g_a$ or $g$, a constraint for their magnitude has to be estimated by including the counter-rotating interactions into the original Hamiltonian $H'$.

Using the James' method~\cite{James,Floquettheory}, the ideal effective Hamiltonian in Eq.~(\ref{Heff}) becomes
\begin{equation}\label{Heffcounter}
\begin{aligned}
H'_{\rm eff}&=\sigma_z\Big[a^\dagger\sum^2_{k=1}g'_ke^{if\sin\phi_k}m_k\\
&-ig'_{12}e^{if(\sin\phi_2-\sin\phi_1)}m^\dag_1m_2+{\rm H.c.}\Big],\\
&+\sigma_z\delta(2\omega_a-n'\omega)\Big[a^\dagger\sum^2_{k=1}G'_ke^{-if\sin\phi_k}m^\dag_k\\
&+G'_{12}e^{-if(\sin\phi_2+\sin\phi_1)}m^\dag_1m^\dag_2+{\rm H.c.}\Big]
\end{aligned}
\end{equation}
with
\begin{equation}\label{countergeff}
\begin{aligned}
g'_k&=-g_agJ_1(f)\cos\phi_k\left(\frac{1}{\omega}-\frac{1}{2\omega_a+\omega}\right), \quad k=1,2, \\
g'_{12}&=2g^2\sum^{\infty}_{n=1}\left(\frac{1}{n\omega}-\frac{1}{2\omega_a+n\omega}\right)
J^2_n(f)\sin\left[n(\phi_2-\phi_1)\right],\\
G'_k&=(-1)^{n'-1}\frac{g_ag}{\omega}\Big[\frac{n'e^{-i(n'-1)\phi_k}J_{n'-1}(f)}{n'-1}\\
&-\frac{n'e^{-i(n'+1)\phi_k}J_{n'+1}(f)}{n'+1}\Big],\\
G'_{12}&=\frac{g^2}{\omega}\sum_{n=1}^{\infty}(-1)^{n+n'}\frac{n'}{n(n'+n)}J_{n}(f)J_{n+n'}(f)\\
&\times\left\{e^{-i\left[n\phi_1-(n+n')\phi_2\right]}+e^{-i\left[n\phi_2-(n+n')\phi_1\right]}\right\}\\
&+\frac{g^2}{\omega}(-1)^{n'}\sum_{n=1}^{n'-1}\frac{1}{n}J_{n}(f)J_{n'-n}(f)\\
&\times\left[e^{in\phi_1+i(n'-n)\phi_2}+e^{in\phi_2+i(n'-n)\phi_1}\right].
\end{aligned}
\end{equation}
The counter rotating interactions ($a^\dag m^\dag_1$, $a^\dag m^\dag_2$, and $m^\dag_1m^\dag_2$) could be omitted when $\omega_a\gg\omega$ (even when $2\omega_a\approx n'\omega$). In this case, the effective coupling strengths $g'_k\approx g_k\gg G_k'$, $g'_{12}\approx g_{12}\gg G'_{12}$, and $J_{n'-1}(f)\ll J_1(f)$, $J_{n}(f)J_{n+n'}(f)\ll J^2_{n}(f)$ and $J_{n}(f)J_{n'-n}(f)\ll J^2_{n}(f)$ for $f=2.4048$ and $n'\gg 1$. On the other hand, the perturbative method based on the James' effective Hamiltonian is valid when the Floquet-driving frequency $\omega\gg g$. Thus, the Floquet driving frequency in our protocol should be in a compromise regime, given transition frequency $\omega_a$ and coupling strength $g$.

\begin{figure}[htbp]
\centering
\includegraphics[width=0.45\textwidth]{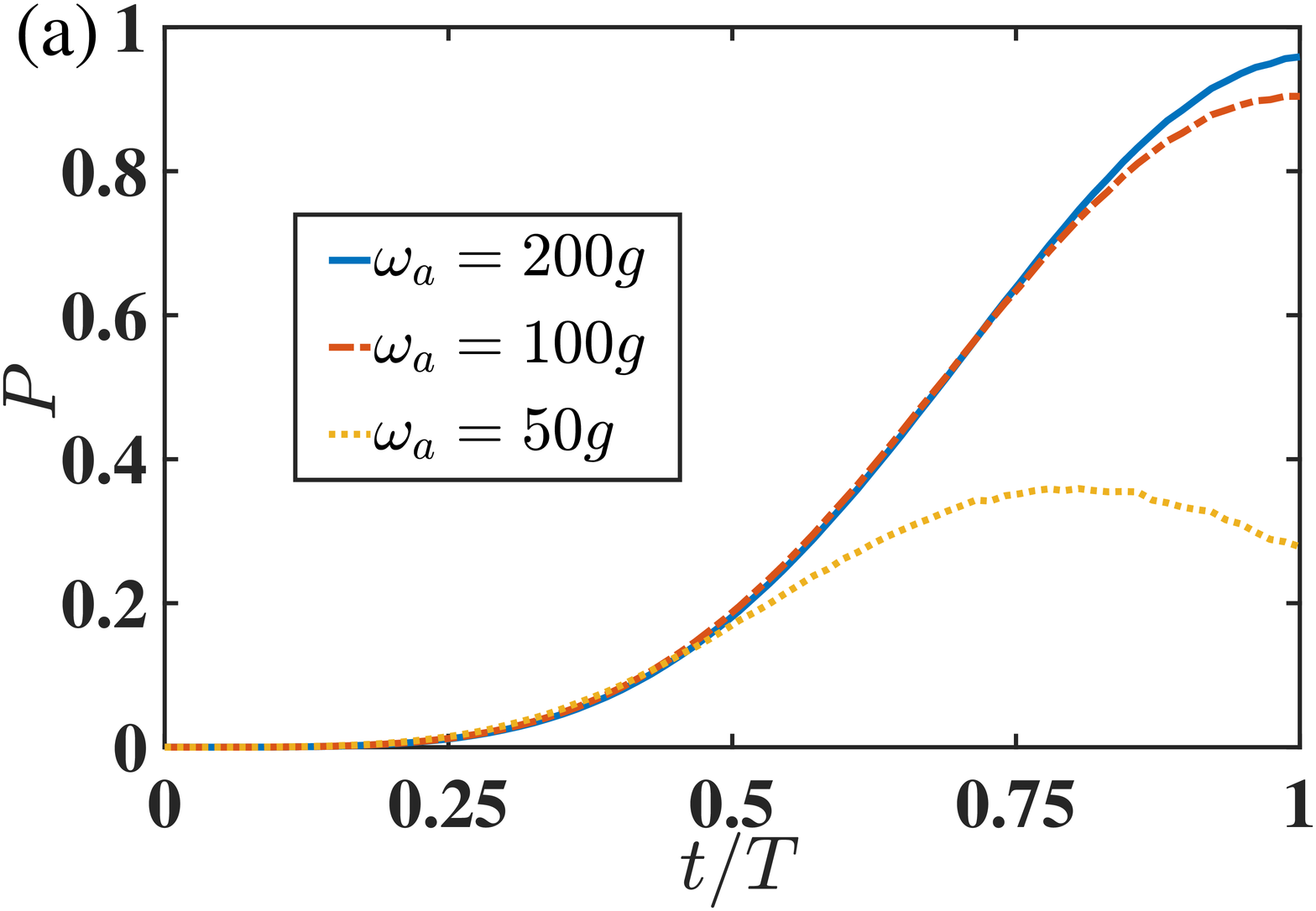}
\includegraphics[width=0.45\textwidth]{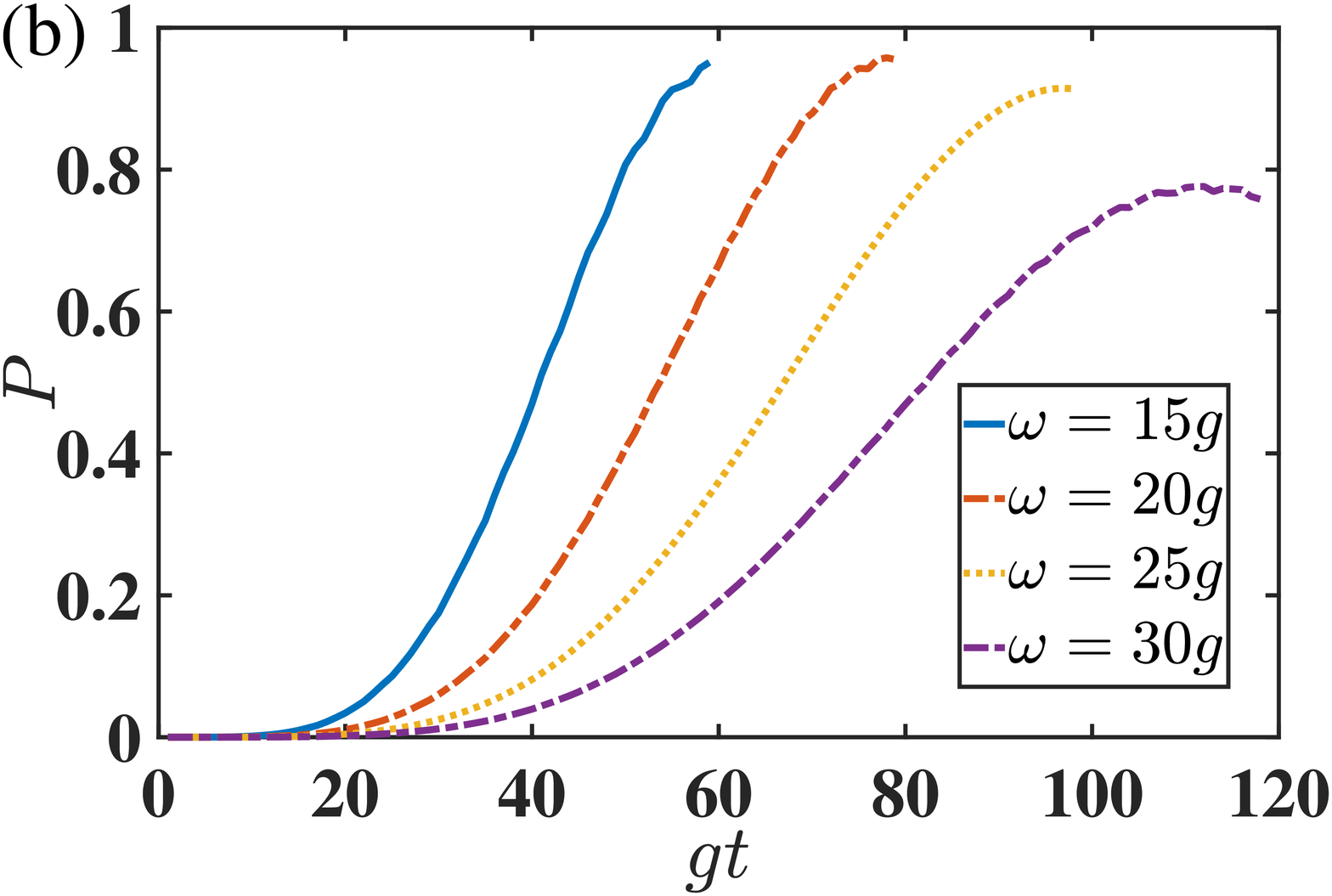}
\caption{Chiral state-transfer fidelity of the target state $(|e002\rangle+|g020\rangle)/\sqrt{2}$ under the system Hamiltonian~(\ref{Hamcounter}) holding the counter-rotating interactions with (a) various transition-frequency $\omega_a$ and a fixed $\omega=20g$, and (b) various Floquet driving frequency $\omega$ and a fixed $\omega_a=200g$. Note the evolution period $T=2\pi/(3\sqrt{3}g_{\rm eff})$ is independent on $\omega_a$ but dependent on $\omega$. The other parameters are set the same as Fig.~\ref{result}.}\label{counter}
\end{figure}

Figure~\ref{counter}(a) is used to reveal the implicitly required magnitude of $\omega_a$ to justify the system Hamiltonian $H$ in our protocol. We demonstrate the effect of the counter-rotating interactions by the chiral state-transfer population dynamics from the initial state $|\Phi(0)\rangle=(|e200\rangle+|g200\rangle)/\sqrt{2}$ under various transition frequency $\omega_a$. It is found that the population can be maintained above $0.90$ when $\omega_a=100g$ and above $0.98$ when $\omega_a=200g$. A smaller $\omega_a$ yields a lower $P$. When $\omega_a$ decreases to $\omega_a=50g$, $P$ is lower than $0.30$ in the end of the evolution. In Fig.~\ref{counter}(b), we plot the chiral state-transfer population dynamics for the same state under various transition frequency $\omega$. It is found that the population can be maintained above $0.90$ in the regime of $15<\omega/g<25$. However, when $\omega=30g$, $P$ is lower than $0.75$ in the end of the evolution with $T=2\pi/(3\sqrt{3}g_{\rm eff})$. Thus our protocol is not appropriate in the strong Floquet regime.

\section{NOON state generation and fidelity analysis}\label{NOONprepar}

This section provides the details of generating the magnonic NOON states based on the Floquet-engineering Hamiltonian~(\ref{Hamiltonian}) under the qubit-dependent chiral-transfer condition. Also we analyze the protocol fidelity in the presence of the quantum dissipation.

\subsection{Preparing NOON state with multiple excitations}\label{prepar}

Suppose that all the four components in our hybrid system (qubit, resonator mode, and two magnonic modes) are initially in their ground states, i.e., $|\Phi(0)\rangle=|g000\rangle=|g\rangle_q|0\rangle_a|0\rangle_{m_1}|0\rangle_{m_2}$. The  generation procedure of a magnonic NOON state could be constituted by two parts, $A$ and $B$.

\begin{figure}[htbp]
\centering
\includegraphics[width=0.45\textwidth]{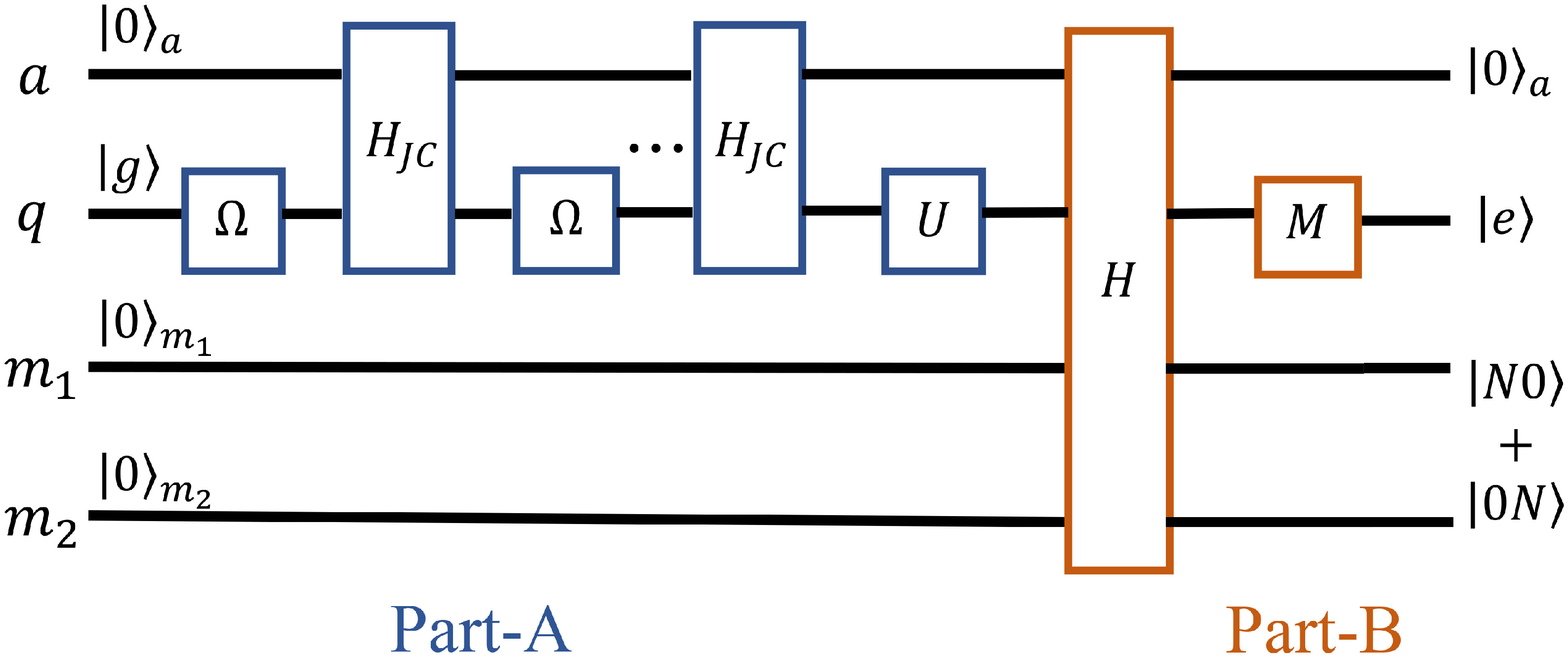}
\caption{Circuit model of the magnonic NOON state generation. Part $A$ is used to prepare the resonator as a Fock state $|N\rangle$ and the qubit as a symmetrical-superposed state. It consists of a sequence of $\pi$-pulses on qubit characterized by the Rabi-frequency $\Omega$, the local evolution of qubit and resonator under a Jaynes-Cummings Hamiltonian $H_{\rm JC}$~(\ref{Hamaq}), and a final $\pi/2$-pulse $U$~(\ref{operatorU}) on qubit. In Part $B$, after the global evolution under the Floquet-engineering Hamiltonian~(\ref{Hamiltonian}), a bare-basis projection is performed on qubit to yield the magnonic NOON state.}\label{model3}
\end{figure}

As illustrated by the circuit model in Fig.~\ref{model3}, Part $A$ is mainly used to prepare the resonator mode as an arbitrary Fock state $|N\rangle$. Accordingly, it is divided into $N+1$ steps as following. This part is local to the resonator and the qubit. Then to avoid the unnecessary crosstalk with the two magnons, the resonator and the qubit are detuned to be far-off-resonant from them. And we temporally remove the time-modulation over the coupling strength between qubit and resonator. Thus in Part A we have a Jaynes-Cummings Hamiltonian
\begin{equation}\label{Hamaq}
H_{\rm JC}=g_a\left(\sigma^+a+\sigma^-a^\dag\right).
\end{equation}

{\em Step} $A_1$. A microwave $\pi$-pulse of $\{\omega_q,\pi/(2\Omega)\}$ is applied to qubit and the system state is transformed to be $|\Phi(\tau)\rangle=|e000\rangle$ up to a global phase. Here $\omega_q$ is the pulse frequency, which is currently resonant with the qubit. $\tau\equiv\pi/(2\Omega)$ is the duration time of the pulse. The Rabi frequency $\Omega$ is assumed to be much larger than $g_a$ to make it reasonable to ignore the evolution under $H_{\rm JC}$ during a sufficiently short period $\tau$. Then turning on the Hamiltonian~(\ref{Hamaq}) for a period of $\tau_1=\pi/(2g_a)$, we have
\begin{equation}\label{state1}
|\Phi(\tau+\tau_1)\rangle=-|g100\rangle.
\end{equation}

{\em Step} $A_j$, $j=2,\cdots,N$. We alternatively employ the same microwave $\pi$-pulse $\{\omega_q,\pi/(2\Omega)\}$ to pump the qubit from the ground state to the excited state and switches on and off the Hamiltonian~(\ref{Hamaq}) with a decreasing duration time $\tau_j=\pi/(2\sqrt{j}g_a)$ to transform $|e(j-1)00\rangle$ to be $|gj00\rangle$. Therefore after these steps, the system state becomes
\begin{equation}\label{state2}
|\Phi(T_A)\rangle=(-1)^N|gN00\rangle,
\end{equation}
where $T_A=N\tau+\sum_{j=1}^N\tau_j$. Till now, a number state is created in the resonator mode. The global phase $(-1)^N$ can be ignored for simplicity.

{\em Step} $A_{N+1}$. A $\pi/2$-pulse gate is performed on qubit. It can be expressed by
\begin{equation}\label{operatorU}
U=e^{i(\pi/4)\vec{\sigma}\cdot\vec{n}}=\frac{1}{\sqrt{2}}
\begin{bmatrix} 1 & ie^{-i\theta}\\ ie^{i\theta} & 1\end{bmatrix},
\end{equation}
where $\vec{\sigma}\equiv(\sigma_x,\sigma_y,\sigma_z)$ and $\vec{n}=(\cos\theta,\sin\theta,0)$. The qubit becomes a superposed state and the system state reads,
\begin{equation}\label{state0}
|\Phi(T_A)\rangle=\frac{1}{\sqrt{2}}\left(|g\rangle+ie^{i\theta}|e\rangle\right)|N00\rangle,
\end{equation}
where the phase $\theta$ is tunable as desired and can be regarded an encoded local phase for the NOON state.

In the beginning of Part $B$, we are well-prepared to generate the magnonic NOON state based on the full Hamiltonian~(\ref{Hamiltonian}) and the system state in Eq.~(\ref{state0}). Here the qubit is in charge of controlling the direction of the chiral state transfer to make the Fock state $|N\rangle$ of the resonator to be perfectly transferred to the two magnonic modes in the same time.

{\em Step} $B_1$. Under the desired time-modulations over the qubit-resonator interaction and the Floquet engineering over the magnonic modes to achieve the effective Hamiltonian in Eq.~(\ref{Heff}), the system state evolves to
\begin{equation}\label{stateT}
|\Phi(T_A+T)\rangle=\frac{1}{\sqrt{2}}\left(|g0N0\rangle-ie^{i\theta}|e00N\rangle\right)
\end{equation}
after $T=2\pi/(3\sqrt{3}g_{\rm eff})$, according to the transformation matrices in Eqs.~(\ref{operatorTe}) and (\ref{operatorTg}). Then we detune the qubit-frequency and apply a microwave $\pi/2$-pulse $\{\omega_q,\pi/(4\Omega)\}$ to the qubit. The state becomes
\begin{equation}\label{stateT}
\begin{aligned}
&|\Phi(T_B=T_A+T+\tau/2)\rangle\\ =&\frac{1}{2}(|g0N0\rangle-e^{i\theta}|g00N\rangle)
-\frac{i}{2}(|e0N0\rangle+e^{i\theta}|e00N\rangle)\\
=&\frac{1}{2}|g0\rangle(|N0\rangle-e^{i\theta}|0N\rangle)
-\frac{i}{2}|e0\rangle(|N0\rangle+e^{i\theta}|0N\rangle).
\end{aligned}
\end{equation}

{\em Step} $B_2$. In the final step, one can obtain the magnonic NOON state $(|N0\rangle-e^{i\theta}|0N\rangle)/\sqrt{2}$ or $(|N0\rangle+e^{i\theta}|0N\rangle)/\sqrt{2}$ by performing a projective measurement on the ground or the excited state of the qubit. To hold the NOON state, the frequencies of all the four components can then be off-set to avoid unnecessary evolution. With a typical coupling strength between the qubit and magnon $g/2\pi\sim20$ MHz~\cite{magnonqubit}, it is found that $T_B\approx 0.65 \mu$s for $N=5$ when the tuning time is omitted (e.g., the pulse duration is nearly $10$ ns~\cite{tunequbit}). So that the full generation time is much smaller than the relaxation time of magnon about $10 \mu$s~\cite{magnon3}.

In a general situation where the resonator is initialled as an arbitrary state $|\varphi\rangle$ and the two magnons are in the same state $|\beta\rangle$, one can generate a (non-normalized) entangled state $|\varphi\rangle|\beta\rangle+|\beta\rangle|\varphi\rangle$ of the two magnonic modes through Part $B$. For example, if $|\varphi\rangle$ is a coherent state $|\alpha\rangle$ and $|\beta\rangle=|0\rangle$, then the final state of two magnonic modes would be a (non-normalized) cat state~\cite{entanglecoher} $|\alpha\rangle|0\rangle+|0\rangle|\alpha\rangle$ as long as the qubit is prepared as the superposed state in Eq.~(\ref{state0}). The state of the magnons could be detected in experiments by the integrated superconducting qubit and NV center~\cite{magnon3,magnonqubit2}.

\subsection{Fidelity Analysis}\label{fideana}

\begin{figure}[htbp]
\centering
\includegraphics[width=0.45\textwidth]{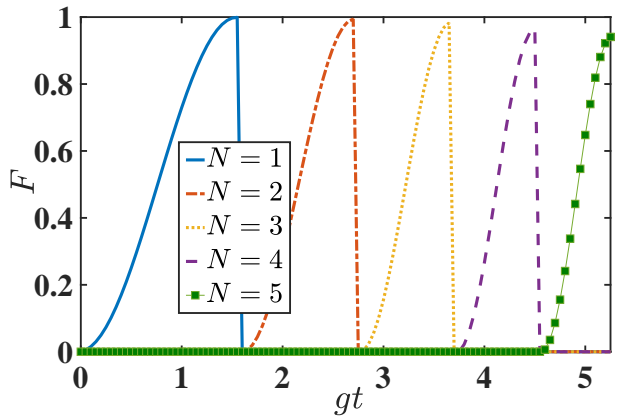}
\caption{Fidelity during the process of generating Fock-state $|N=5\rangle$ of the resonator mode $a$ as indicated in Eq.~(\ref{state2}). All the lines except the green solid line marked with squares indicate the intermediate result about the overlap between evolved state and $|N<5\rangle$. $\Omega=30g$, $g_a=g$, and $\gamma/g=10^{-5}$.}\label{Fockdynamics}
\end{figure}

The physical feasibility of our generation protocol could be verified by the numerical simulation over the whole procedure in Sec.~\ref{prepar}. The overlap of the final state and the ideal NOON state in the end of {\em Step} $B_2$ could be regarded as the generation-protocol fidelity. In particular, we prepare the whole system as a product of ground state, follow all the steps in Parts $A$ and $B$, and take account the decoherence from all components of the hybrid system into the dynamics. Under the standard Markovian approximation and tracing out the degrees of freedom of the external environment (assumed to be at the vacuum state), we arrive at the master equation for the density-matrix operator $\rho(t)$ of the whole system consisting of qubit, resonator, and magnonic modes:
\begin{equation}\label{master}
\begin{aligned}
\dot{\rho}(t)&=-i\left[\tilde{H}, \rho(t)\right]+\kappa_a\mathcal{L}[a]\rho(t)+\kappa_m\mathcal{L}[m_1]\rho(t) \\
&+\kappa_m\mathcal{L}[m_2]\rho(t)+\gamma\mathcal{L}[\sigma^-]\rho(t).
\end{aligned}
\end{equation}
In Part $A$ for the local evolution of qubit and resonator, $\tilde{H}=H_{\rm JC}$~(\ref{Hamaq}); and in Part $B$ for the global evolution, $\tilde{H}=H$~(\ref{Hamiltonian}). $\kappa_a$, $\kappa_m$, and $\gamma$ are the decoherence rates of the resonator, the magnonic modes, and the qubit, respectively. To simplify the discussion but with no loss of generality, we set $\kappa_a=\kappa_m=\gamma$. These rates are surely dependent on the magnitude of the associated transitions in spontaneous emission. Yet this setting is simply used to estimate the robustness of the ideal protocol, so all of the decoherence rates are supposed to be in the same order of magnitude. The dissipative superoperator $\mathcal{L}$ is defined in a Lindblad form,
\begin{equation}\label{operator}
\mathcal{L}[o]\rho\equiv\frac{1}{2}\left(2o\rho o^\dag-o^\dag o\rho-\rho o^\dag o\right),
\end{equation}
where $o=\sigma^-,a,m_1,m_2$ are the decay operators.

Before taking the crucial steps of generating NOON state by chiral state transfer, as discussed in Sec.~\ref{secmodel}, we have to consider the nonideal dynamics during Part $A$ to achieve a desired Fock state of the resonator. In Fig.~\ref{Fockdynamics}, the fidelity dynamics of $|N=5\rangle$ is plotted by using the master equation~(\ref{master}). On this stage, the magnons are separable. The five peak values indicates the temporal fidelity of $|N\leq5\rangle$ during the step-by-step process, which decline with $N$. In particular, we have $F=0.95$ for $N=5$.

\begin{figure}[htbp]
\centering
\includegraphics[width=0.45\textwidth]{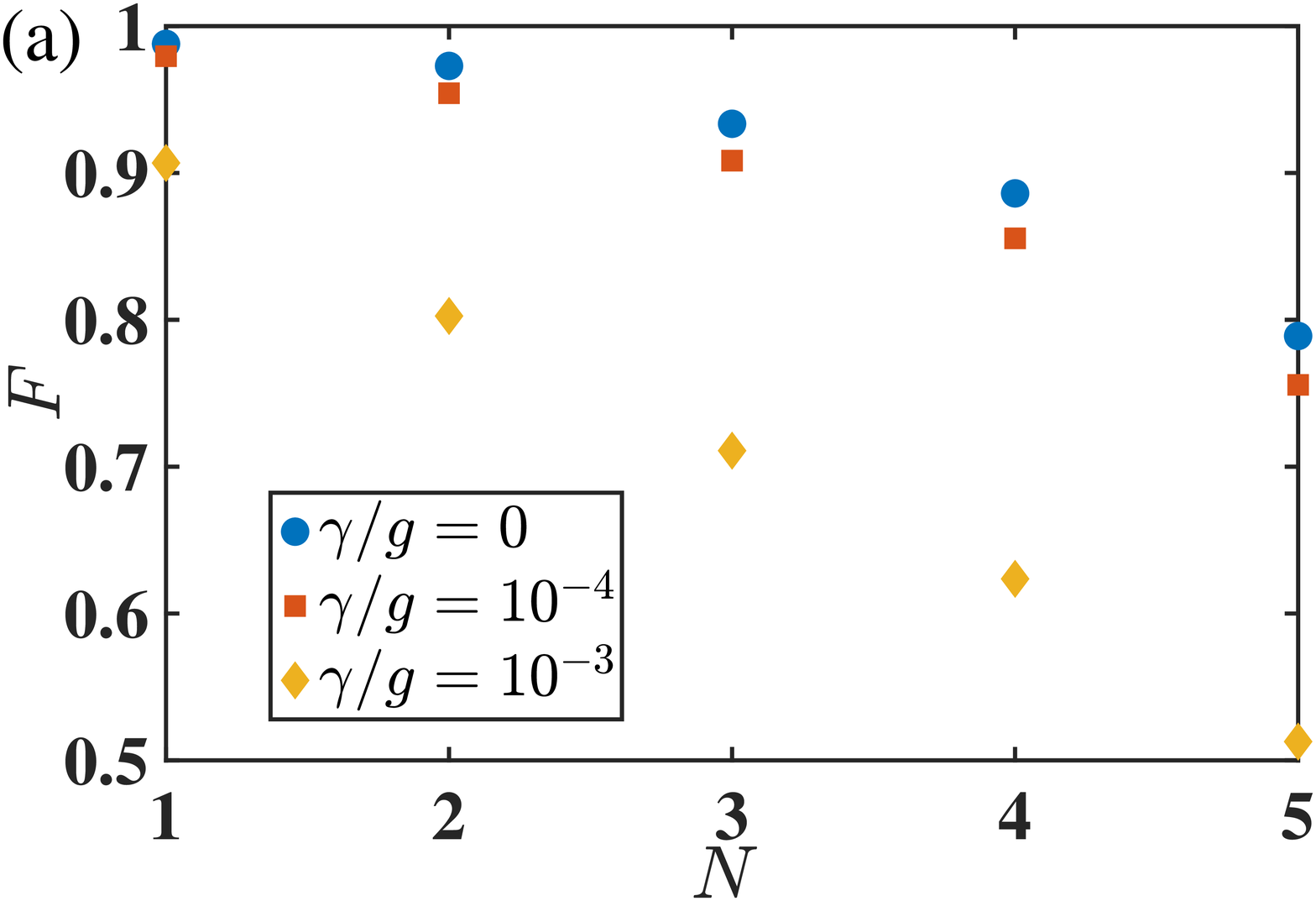}
\includegraphics[width=0.45\textwidth]{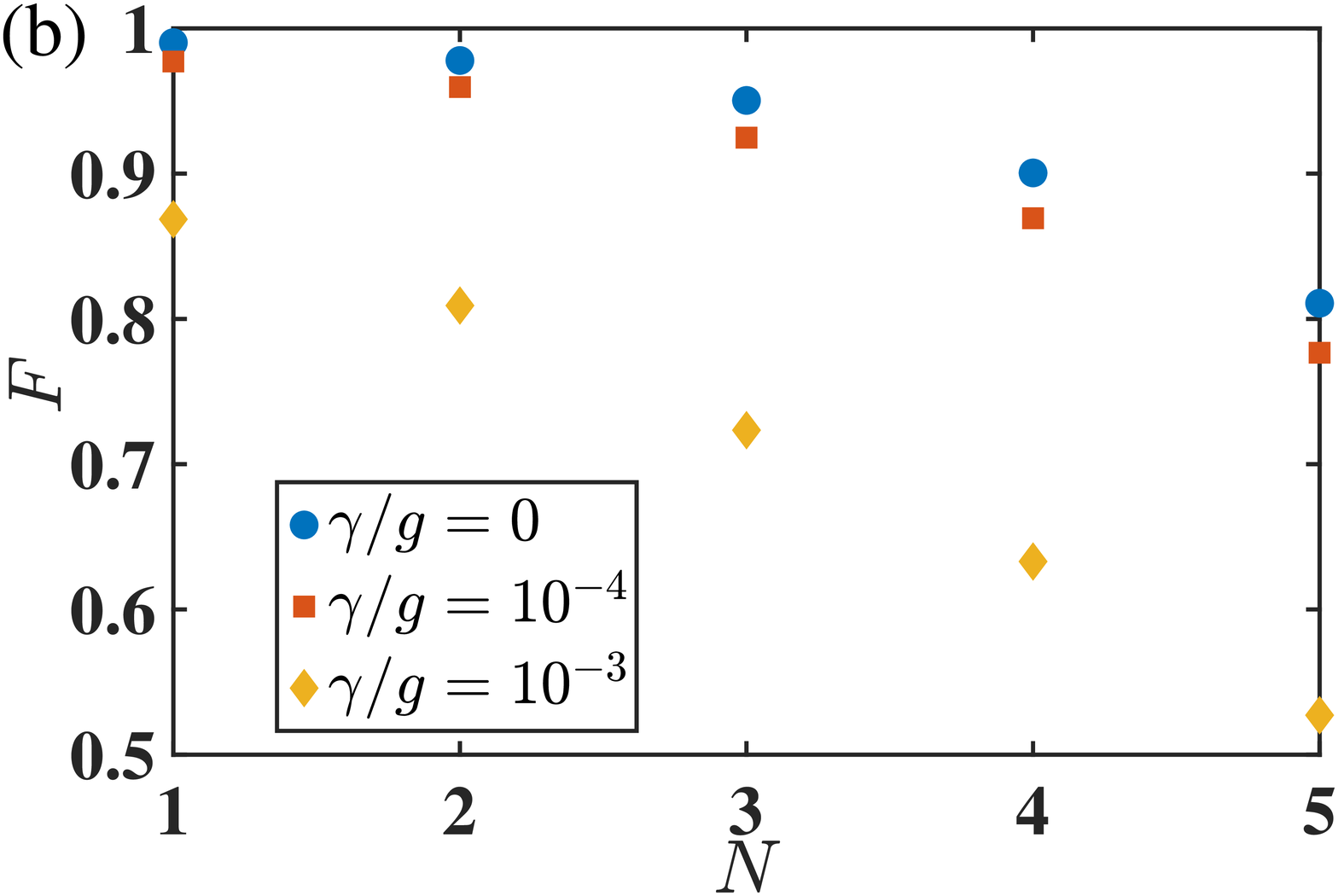}
\caption{Fidelity of the NOON state obtained by the projective measurement over (a) the excited state and (b) the ground state of the qubit (see {\em Step} $B_2$ in Sec.~\ref{prepar}), under various decoherence rate $\gamma$. Here the parameters are set as $\omega=20g$, $\phi_1=2\pi/3$, $\phi_2=4\pi/3$, $\theta=\pi$, and $\Omega=30g$.}\label{masterfig}
\end{figure}

Using Eq.~(\ref{stateT}), {\em Step} $B_2$, and Eq.~(\ref{master}), the final fidelity is calculated by $F=\langle\Psi|{\rm Tr}_{\rm qr}[\rho(T_B)]|\Psi\rangle$, where $|\Psi\rangle$ is the target magnonic NOON state and ${\rm Tr}_{\rm qr}[\cdot]$ means tracing out the degrees of freedom of qubit and resonator. Performing $|e\rangle\langle e|$ and $|g\rangle\langle g|$ on $|\Phi(T_B)\rangle$ gives rise to $|\Psi\rangle=(|N0\rangle+e^{i\theta}|0N\rangle)/\sqrt{2}$ and $|\Psi\rangle=(|N0\rangle-e^{i\theta}|0N\rangle)/\sqrt{2}$, respectively. In Fig.~\ref{masterfig}, we plot the fidelity of the NOON state under various decoherence rates $\gamma$. As expected, the fidelities decline with $N$. In comparison of Figs.~\ref{masterfig}(a) and \ref{masterfig}(b), the results are not sensitive to the choice of the measurement basis. With a moderate decoherence rate $\gamma/g=10^{-4}$, the fidelities are $97.9\%$, $95.4\%$, $90.1\%$, $85.6\%$, and $75.6\%$ for $N=1,2,3,4,5$ in Fig.~\ref{masterfig}(a) by projecting to the excited state $|e\rangle$. And they are respectively $97.8\%$, $96.0\%$, $92.4\%$, $86.9\%$, and $77.7\%$ in Fig.~\ref{masterfig}(b) by measuring the ground state. Roughly the latter are slightly higher than the former and both demonstrate robustness against the environmental dissipation. Under a strong decoherence rate $\gamma/g=10^{-3}$, however, the generation fidelity will decline to almost $50\%$.

\section{Conclusion}\label{conclu}

In summary, we have presented a magnonic NOON state generation protocol based on the Floquet engineering method. The control protocol is carried out in a hybrid qubit-resonator-magnon system, where the qubit is coupled to the resonator mode with a time-modulation interaction and simultaneously coupled to two magnonic modes. The Floquet engineering with desired driving intensity, frequency, and local phases is applied to the eigen-frequencies of two magnons, by which an effective time-reversal-symmetry broken Hamiltonian is constructed to realize a chiral state transfer amongst the resonator and the two magnonic modes. The state transfer direction could be controlled by the state of the qubit. In our protocol, when the qubit is prepared as $(|e\rangle+|g\rangle)/\sqrt{2}$, an arbitrary pure state (including the Fock state) of the resonator can thus be transferred to the two magnons at the same time. We can eventually obtain the magnonic NOON state upon a projective measurement on qubit. We estimate the sensitivity of the state-transfer fidelity to the systematic errors in qubit-magnon coupling strength, Floquet driving intensity, frequency mismatch in magnons and qubit, and counter-rotating interactions. Our protocol shows robustness against the quantum dissipation of all the components. Our work therefore provides an alternative approach to generate NOON state in a novel hybrid system, which constitutes an interesting application of chiral state transfer under Floquet control.

\section*{Acknowledgments}

We acknowledge financial support from the National Science Foundation of China (Grants No. 11974311 and No. U1801661).

\bibliographystyle{apsrevlong}
\bibliography{reference}

\begin{thebibliography}{60}%
\makeatletter
\providecommand \@ifxundefined [1]{%
 \@ifx{#1\undefined}
}%
\providecommand \@ifnum [1]{%
 \ifnum #1\expandafter \@firstoftwo
 \else \expandafter \@secondoftwo
 \fi
}%
\providecommand \@ifx [1]{%
 \ifx #1\expandafter \@firstoftwo
 \else \expandafter \@secondoftwo
 \fi
}%
\providecommand \natexlab [1]{#1}%
\providecommand \enquote  [1]{``#1''}%
\providecommand \bibnamefont  [1]{#1}%
\providecommand \bibfnamefont [1]{#1}%
\providecommand \citenamefont [1]{#1}%
\providecommand \href@noop [0]{\@secondoftwo}%
\providecommand \href [0]{\begingroup \@sanitize@url \@href}%
\providecommand \@href[1]{\@@startlink{#1}\@@href}%
\providecommand \@@href[1]{\endgroup#1\@@endlink}%
\providecommand \@sanitize@url [0]{\catcode `\\12\catcode `\$12\catcode
  `\&12\catcode `\#12\catcode `\^12\catcode `\_12\catcode `\%12\relax}%
\providecommand \@@startlink[1]{}%
\providecommand \@@endlink[0]{}%
\providecommand \url  [0]{\begingroup\@sanitize@url \@url }%
\providecommand \@url [1]{\endgroup\@href {#1}{\urlprefix }}%
\providecommand \urlprefix  [0]{URL }%
\providecommand \Eprint [0]{\href }%
\providecommand \doibase [0]{http://dx.doi.org/}%
\providecommand \selectlanguage [0]{\@gobble}%
\providecommand \bibinfo  [0]{\@secondoftwo}%
\providecommand \bibfield  [0]{\@secondoftwo}%
\providecommand \translation [1]{[#1]}%
\providecommand \BibitemOpen [0]{}%
\providecommand \bibitemStop [0]{}%
\providecommand \bibitemNoStop [0]{.\EOS\space}%
\providecommand \EOS [0]{\spacefactor3000\relax}%
\providecommand \BibitemShut  [1]{\csname bibitem#1\endcsname}%
\let\auto@bib@innerbib\@empty
\bibitem [{\citenamefont {Boto}\ \emph {et~al.}(2000)\citenamefont {Boto},
  \citenamefont {Kok}, \citenamefont {Abrams}, \citenamefont {Braunstein},
  \citenamefont {Williams},\ and\ \citenamefont {Dowling}}]{quaninter}%
  \BibitemOpen
  \bibfield  {author} {\bibinfo {author} {\bibfnamefont {A.~N.}\ \bibnamefont
  {Boto}}, \bibinfo {author} {\bibfnamefont {P.}~\bibnamefont {Kok}}, \bibinfo
  {author} {\bibfnamefont {D.~S.}\ \bibnamefont {Abrams}}, \bibinfo {author}
  {\bibfnamefont {S.~L.}\ \bibnamefont {Braunstein}}, \bibinfo {author}
  {\bibfnamefont {C.~P.}\ \bibnamefont {Williams}}, \ and\ \bibinfo {author}
  {\bibfnamefont {J.~P.}\ \bibnamefont {Dowling}},\ }\bibfield  {title} {\emph
  {\bibinfo {title} {Quantum interferometric optical lithography: Exploiting
  entanglement to beat the diffraction limit},\ }}\href {\doibase
  10.1103/PhysRevLett.85.2733} {\bibfield  {journal} {\bibinfo  {journal}
  {Phys. Rev. Lett.}\ }\textbf {\bibinfo {volume} {85}},\ \bibinfo {pages}
  {2733} (\bibinfo {year} {2000})}\BibitemShut {NoStop}%
\bibitem [{\citenamefont {Horodecki}\ \emph {et~al.}(2009)\citenamefont
  {Horodecki}, \citenamefont {Horodecki}, \citenamefont {Horodecki},\ and\
  \citenamefont {Horodecki}}]{qe}%
  \BibitemOpen
  \bibfield  {author} {\bibinfo {author} {\bibfnamefont {R.}~\bibnamefont
  {Horodecki}}, \bibinfo {author} {\bibfnamefont {P.}~\bibnamefont
  {Horodecki}}, \bibinfo {author} {\bibfnamefont {M.}~\bibnamefont
  {Horodecki}}, \ and\ \bibinfo {author} {\bibfnamefont {K.}~\bibnamefont
  {Horodecki}},\ }\bibfield  {title} {\emph {\bibinfo {title} {Quantum
  entanglement},\ }}\href {\doibase 10.1103/RevModPhys.81.865} {\bibfield
  {journal} {\bibinfo  {journal} {Rev. Mod. Phys.}\ }\textbf {\bibinfo {volume}
  {81}},\ \bibinfo {pages} {865} (\bibinfo {year} {2009})}\BibitemShut
  {NoStop}%
\bibitem [{\citenamefont {Pezz\`e}\ \emph {et~al.}(2018)\citenamefont
  {Pezz\`e}, \citenamefont {Smerzi}, \citenamefont {Oberthaler}, \citenamefont
  {Schmied},\ and\ \citenamefont {Treutlein}}]{quantmetro}%
  \BibitemOpen
  \bibfield  {author} {\bibinfo {author} {\bibfnamefont {L.}~\bibnamefont
  {Pezz\`e}}, \bibinfo {author} {\bibfnamefont {A.}~\bibnamefont {Smerzi}},
  \bibinfo {author} {\bibfnamefont {M.~K.}\ \bibnamefont {Oberthaler}},
  \bibinfo {author} {\bibfnamefont {R.}~\bibnamefont {Schmied}}, \ and\
  \bibinfo {author} {\bibfnamefont {P.}~\bibnamefont {Treutlein}},\ }\bibfield
  {title} {\emph {\bibinfo {title} {Quantum metrology with nonclassical states
  of atomic ensembles},\ }}\href {\doibase 10.1103/RevModPhys.90.035005}
  {\bibfield  {journal} {\bibinfo  {journal} {Rev. Mod. Phys.}\ }\textbf
  {\bibinfo {volume} {90}},\ \bibinfo {pages} {035005} (\bibinfo {year}
  {2018})}\BibitemShut {NoStop}%
\bibitem [{\citenamefont {Gisin}\ and\ \citenamefont {Thew}(2007)}]{qco}%
  \BibitemOpen
  \bibfield  {author} {\bibinfo {author} {\bibfnamefont {N.}~\bibnamefont
  {Gisin}}\ and\ \bibinfo {author} {\bibfnamefont {R.}~\bibnamefont {Thew}},\
  }\bibfield  {title} {\emph {\bibinfo {title} {Quantum communication},\
  }}\href@noop {} {\bibfield  {journal} {\bibinfo  {journal} {Nat. Photon.}\
  }\textbf {\bibinfo {volume} {1}},\ \bibinfo {pages} {165} (\bibinfo {year}
  {2007})}\BibitemShut {NoStop}%
\bibitem [{\citenamefont {Bennett}\ and\ \citenamefont
  {DiVincenzo}(2000)}]{quantinf}%
  \BibitemOpen
  \bibfield  {author} {\bibinfo {author} {\bibfnamefont {C.}~\bibnamefont
  {Bennett}}\ and\ \bibinfo {author} {\bibfnamefont {B.}~\bibnamefont
  {DiVincenzo}},\ }\bibfield  {title} {\emph {\bibinfo {title} {Quantum
  information and computation},\ }}\href {\doibase 10.1038/35005001} {\bibfield
   {journal} {\bibinfo  {journal} {Nature}\ }\textbf {\bibinfo {volume}
  {404}},\ \bibinfo {pages} {247} (\bibinfo {year} {2000})}\BibitemShut
  {NoStop}%
\bibitem [{\citenamefont {Strauch}\ \emph {et~al.}(2010)\citenamefont
  {Strauch}, \citenamefont {Jacobs},\ and\ \citenamefont {Simmonds}}]{noon1}%
  \BibitemOpen
  \bibfield  {author} {\bibinfo {author} {\bibfnamefont {F.~W.}\ \bibnamefont
  {Strauch}}, \bibinfo {author} {\bibfnamefont {K.}~\bibnamefont {Jacobs}}, \
  and\ \bibinfo {author} {\bibfnamefont {R.~W.}\ \bibnamefont {Simmonds}},\
  }\bibfield  {title} {\emph {\bibinfo {title} {Arbitrary control of
  entanglement between two superconducting resonators},\ }}\href {\doibase
  10.1103/PhysRevLett.105.050501} {\bibfield  {journal} {\bibinfo  {journal}
  {Phys. Rev. Lett.}\ }\textbf {\bibinfo {volume} {105}},\ \bibinfo {pages}
  {050501} (\bibinfo {year} {2010})}\BibitemShut {NoStop}%
\bibitem [{\citenamefont {Merkel}\ and\ \citenamefont {Wilhelm}(2010)}]{noon2}%
  \BibitemOpen
  \bibfield  {author} {\bibinfo {author} {\bibfnamefont {S.~T.}\ \bibnamefont
  {Merkel}}\ and\ \bibinfo {author} {\bibfnamefont {F.~K.}\ \bibnamefont
  {Wilhelm}},\ }\bibfield  {title} {\emph {\bibinfo {title} {Generation and
  detection of noon states in superconducting circuits},\ }}\href@noop {}
  {\bibfield  {journal} {\bibinfo  {journal} {New J. Phys.}\ }\textbf {\bibinfo
  {volume} {12}},\ \bibinfo {pages} {3175} (\bibinfo {year}
  {2010})}\BibitemShut {NoStop}%
\bibitem [{\citenamefont {Wang}\ \emph {et~al.}(2011)\citenamefont {Wang},
  \citenamefont {Mariantoni}, \citenamefont {Bialczak}, \citenamefont
  {Lenander}, \citenamefont {Lucero}, \citenamefont {Neeley}, \citenamefont
  {O'Connell}, \citenamefont {Sank}, \citenamefont {Weides}, \citenamefont
  {Wenner}, \citenamefont {Yamamoto}, \citenamefont {Yin}, \citenamefont
  {Zhao}, \citenamefont {Martinis},\ and\ \citenamefont {Cleland}}]{noon3}%
  \BibitemOpen
  \bibfield  {author} {\bibinfo {author} {\bibfnamefont {H.}~\bibnamefont
  {Wang}}, \bibinfo {author} {\bibfnamefont {M.}~\bibnamefont {Mariantoni}},
  \bibinfo {author} {\bibfnamefont {R.~C.}\ \bibnamefont {Bialczak}}, \bibinfo
  {author} {\bibfnamefont {M.}~\bibnamefont {Lenander}}, \bibinfo {author}
  {\bibfnamefont {E.}~\bibnamefont {Lucero}}, \bibinfo {author} {\bibfnamefont
  {M.}~\bibnamefont {Neeley}}, \bibinfo {author} {\bibfnamefont {A.~D.}\
  \bibnamefont {O'Connell}}, \bibinfo {author} {\bibfnamefont {D.}~\bibnamefont
  {Sank}}, \bibinfo {author} {\bibfnamefont {M.}~\bibnamefont {Weides}},
  \bibinfo {author} {\bibfnamefont {J.}~\bibnamefont {Wenner}}, \bibinfo
  {author} {\bibfnamefont {T.}~\bibnamefont {Yamamoto}}, \bibinfo {author}
  {\bibfnamefont {Y.}~\bibnamefont {Yin}}, \bibinfo {author} {\bibfnamefont
  {J.}~\bibnamefont {Zhao}}, \bibinfo {author} {\bibfnamefont {J.~M.}\
  \bibnamefont {Martinis}}, \ and\ \bibinfo {author} {\bibfnamefont {A.~N.}\
  \bibnamefont {Cleland}},\ }\bibfield  {title} {\emph {\bibinfo {title}
  {Deterministic entanglement of photons in two superconducting microwave
  resonators},\ }}\href {\doibase 10.1103/PhysRevLett.106.060401} {\bibfield
  {journal} {\bibinfo  {journal} {Phys. Rev. Lett.}\ }\textbf {\bibinfo
  {volume} {106}},\ \bibinfo {pages} {060401} (\bibinfo {year}
  {2011})}\BibitemShut {NoStop}%
\bibitem [{\citenamefont {Su}\ \emph {et~al.}(2014)\citenamefont {Su},
  \citenamefont {Yang},\ and\ \citenamefont {Zheng}}]{noon4}%
  \BibitemOpen
  \bibfield  {author} {\bibinfo {author} {\bibfnamefont {Q.~P.}\ \bibnamefont
  {Su}}, \bibinfo {author} {\bibfnamefont {C.~P.}\ \bibnamefont {Yang}}, \ and\
  \bibinfo {author} {\bibfnamefont {S.~B.}\ \bibnamefont {Zheng}},\ }\bibfield
  {title} {\emph {\bibinfo {title} {Fast and simple scheme for generating noon
  states of photons in circuit qed},\ }}\href@noop {} {\bibfield  {journal}
  {\bibinfo  {journal} {Sci. Rep.}\ }\textbf {\bibinfo {volume} {4}},\ \bibinfo
  {pages} {3898} (\bibinfo {year} {2014})}\BibitemShut {NoStop}%
\bibitem [{\citenamefont {Xiong}\ \emph {et~al.}(2015)\citenamefont {Xiong},
  \citenamefont {Sun}, \citenamefont {Liu}, \citenamefont {Liu},\ and\
  \citenamefont {Yang}}]{noon5}%
  \BibitemOpen
  \bibfield  {author} {\bibinfo {author} {\bibfnamefont {S.~J.}\ \bibnamefont
  {Xiong}}, \bibinfo {author} {\bibfnamefont {Z.}~\bibnamefont {Sun}}, \bibinfo
  {author} {\bibfnamefont {J.~M.}\ \bibnamefont {Liu}}, \bibinfo {author}
  {\bibfnamefont {T.}~\bibnamefont {Liu}}, \ and\ \bibinfo {author}
  {\bibfnamefont {C.~P.}\ \bibnamefont {Yang}},\ }\bibfield  {title} {\emph
  {\bibinfo {title} {Efficient scheme for generation of photonic noon states in
  circuit qed},\ }}\href@noop {} {\bibfield  {journal} {\bibinfo  {journal}
  {Opt. Lett.}\ }\textbf {\bibinfo {volume} {40}},\ \bibinfo {pages} {2221}
  (\bibinfo {year} {2015})}\BibitemShut {NoStop}%
\bibitem [{\citenamefont {Su}\ \emph {et~al.}(2017)\citenamefont {Su},
  \citenamefont {Zhu}, \citenamefont {Yu}, \citenamefont {Zhang}, \citenamefont
  {Xiong}, \citenamefont {Liu},\ and\ \citenamefont {Yang}}]{noon6}%
  \BibitemOpen
  \bibfield  {author} {\bibinfo {author} {\bibfnamefont {Q.-P.}\ \bibnamefont
  {Su}}, \bibinfo {author} {\bibfnamefont {H.-H.}\ \bibnamefont {Zhu}},
  \bibinfo {author} {\bibfnamefont {L.}~\bibnamefont {Yu}}, \bibinfo {author}
  {\bibfnamefont {Y.}~\bibnamefont {Zhang}}, \bibinfo {author} {\bibfnamefont
  {S.-J.}\ \bibnamefont {Xiong}}, \bibinfo {author} {\bibfnamefont {J.-M.}\
  \bibnamefont {Liu}}, \ and\ \bibinfo {author} {\bibfnamefont {C.-P.}\
  \bibnamefont {Yang}},\ }\bibfield  {title} {\emph {\bibinfo {title}
  {Generating double noon states of photons in circuit qed},\ }}\href {\doibase
  10.1103/PhysRevA.95.022339} {\bibfield  {journal} {\bibinfo  {journal} {Phys.
  Rev. A}\ }\textbf {\bibinfo {volume} {95}},\ \bibinfo {pages} {022339}
  (\bibinfo {year} {2017})}\BibitemShut {NoStop}%
\bibitem [{\citenamefont {Qi}\ and\ \citenamefont {Jing}(2020)}]{noon7}%
  \BibitemOpen
  \bibfield  {author} {\bibinfo {author} {\bibfnamefont {S.-f.}\ \bibnamefont
  {Qi}}\ and\ \bibinfo {author} {\bibfnamefont {J.}~\bibnamefont {Jing}},\
  }\bibfield  {title} {\emph {\bibinfo {title} {Generating noon states in
  circuit qed using a multiphoton resonance in the presence of counter-rotating
  interactions},\ }}\href {\doibase 10.1103/PhysRevA.101.033809} {\bibfield
  {journal} {\bibinfo  {journal} {Phys. Rev. A}\ }\textbf {\bibinfo {volume}
  {101}},\ \bibinfo {pages} {033809} (\bibinfo {year} {2020})}\BibitemShut
  {NoStop}%
\bibitem [{\citenamefont {D'Angelo}\ \emph {et~al.}(2008)\citenamefont
  {D'Angelo}, \citenamefont {Garuccio},\ and\ \citenamefont {Tamma}}]{noon8}%
  \BibitemOpen
  \bibfield  {author} {\bibinfo {author} {\bibfnamefont {M.}~\bibnamefont
  {D'Angelo}}, \bibinfo {author} {\bibfnamefont {A.}~\bibnamefont {Garuccio}},
  \ and\ \bibinfo {author} {\bibfnamefont {V.}~\bibnamefont {Tamma}},\
  }\bibfield  {title} {\emph {\bibinfo {title} {Toward real maximally
  path-entangled $n$-photon-state sources},\ }}\href {\doibase
  10.1103/PhysRevA.77.063826} {\bibfield  {journal} {\bibinfo  {journal} {Phys.
  Rev. A}\ }\textbf {\bibinfo {volume} {77}},\ \bibinfo {pages} {063826}
  (\bibinfo {year} {2008})}\BibitemShut {NoStop}%
\bibitem [{\citenamefont {Kamide}\ \emph {et~al.}(2017)\citenamefont {Kamide},
  \citenamefont {Ota}, \citenamefont {Iwamoto},\ and\ \citenamefont
  {Arakawa}}]{noon9}%
  \BibitemOpen
  \bibfield  {author} {\bibinfo {author} {\bibfnamefont {K.}~\bibnamefont
  {Kamide}}, \bibinfo {author} {\bibfnamefont {Y.}~\bibnamefont {Ota}},
  \bibinfo {author} {\bibfnamefont {S.}~\bibnamefont {Iwamoto}}, \ and\
  \bibinfo {author} {\bibfnamefont {Y.}~\bibnamefont {Arakawa}},\ }\bibfield
  {title} {\emph {\bibinfo {title} {Method for generating a photonic noon state
  with quantum dots in coupled nanocavities},\ }}\href {\doibase
  10.1103/PhysRevA.96.013853} {\bibfield  {journal} {\bibinfo  {journal} {Phys.
  Rev. A}\ }\textbf {\bibinfo {volume} {96}},\ \bibinfo {pages} {013853}
  (\bibinfo {year} {2017})}\BibitemShut {NoStop}%
\bibitem [{\citenamefont {Vanhaele}\ \emph {et~al.}(2022)\citenamefont
  {Vanhaele}, \citenamefont {B\"acker}, \citenamefont {Ketzmerick},\ and\
  \citenamefont {Schlagheck}}]{triplenoon}%
  \BibitemOpen
  \bibfield  {author} {\bibinfo {author} {\bibfnamefont {G.}~\bibnamefont
  {Vanhaele}}, \bibinfo {author} {\bibfnamefont {A.}~\bibnamefont {B\"acker}},
  \bibinfo {author} {\bibfnamefont {R.}~\bibnamefont {Ketzmerick}}, \ and\
  \bibinfo {author} {\bibfnamefont {P.}~\bibnamefont {Schlagheck}},\ }\bibfield
   {title} {\emph {\bibinfo {title} {Creating triple-noon states with ultracold
  atoms via chaos-assisted tunneling},\ }}\href {\doibase
  10.1103/PhysRevA.106.L011301} {\bibfield  {journal} {\bibinfo  {journal}
  {Phys. Rev. A}\ }\textbf {\bibinfo {volume} {106}},\ \bibinfo {pages}
  {L011301} (\bibinfo {year} {2022})}\BibitemShut {NoStop}%
\bibitem [{\citenamefont {Israel}\ \emph {et~al.}(2014)\citenamefont {Israel},
  \citenamefont {Rosen},\ and\ \citenamefont {Silberberg}}]{polarnoon}%
  \BibitemOpen
  \bibfield  {author} {\bibinfo {author} {\bibfnamefont {Y.}~\bibnamefont
  {Israel}}, \bibinfo {author} {\bibfnamefont {S.}~\bibnamefont {Rosen}}, \
  and\ \bibinfo {author} {\bibfnamefont {Y.}~\bibnamefont {Silberberg}},\
  }\bibfield  {title} {\emph {\bibinfo {title} {Supersensitive polarization
  microscopy using noon states of light},\ }}\href {\doibase
  10.1103/PhysRevLett.112.103604} {\bibfield  {journal} {\bibinfo  {journal}
  {Phys. Rev. Lett.}\ }\textbf {\bibinfo {volume} {112}},\ \bibinfo {pages}
  {103604} (\bibinfo {year} {2014})}\BibitemShut {NoStop}%
\bibitem [{\citenamefont {Jones}\ \emph {et~al.}(2009)\citenamefont {Jones},
  \citenamefont {Karlen}, \citenamefont {Fitzsimons}, \citenamefont {Ardavan},
  \citenamefont {Briggs},\ and\ \citenamefont {Morton}}]{magnetsense}%
  \BibitemOpen
  \bibfield  {author} {\bibinfo {author} {\bibfnamefont {J.~A.}\ \bibnamefont
  {Jones}}, \bibinfo {author} {\bibfnamefont {S.~D.}\ \bibnamefont {Karlen}},
  \bibinfo {author} {\bibfnamefont {J.}~\bibnamefont {Fitzsimons}}, \bibinfo
  {author} {\bibfnamefont {S.~C.}\ \bibnamefont {Ardavan}, \bibfnamefont
  {A.~Benjamin}}, \bibinfo {author} {\bibfnamefont {A.~D.}\ \bibnamefont
  {Briggs}}, \ and\ \bibinfo {author} {\bibfnamefont {J.~J.~L.}\ \bibnamefont
  {Morton}},\ }\bibfield  {title} {\emph {\bibinfo {title} {Magnetic field
  sensing beyond the standard quantum limit using 10-spin noon states},\
  }}\href {\doibase 10.1126/science.1170730} {\bibfield  {journal} {\bibinfo
  {journal} {Science}\ }\textbf {\bibinfo {volume} {324}},\ \bibinfo {pages}
  {1166} (\bibinfo {year} {2009})}\BibitemShut {NoStop}%
\bibitem [{\citenamefont {Afek}\ \emph {et~al.}(2010)\citenamefont {Afek},
  \citenamefont {Ambar},\ and\ \citenamefont {Silberberg}}]{lightnoon}%
  \BibitemOpen
  \bibfield  {author} {\bibinfo {author} {\bibfnamefont {I.}~\bibnamefont
  {Afek}}, \bibinfo {author} {\bibfnamefont {O.}~\bibnamefont {Ambar}}, \ and\
  \bibinfo {author} {\bibfnamefont {Y.}~\bibnamefont {Silberberg}},\ }\bibfield
   {title} {\emph {\bibinfo {title} {High-noon states by mixing quantum and
  classical light},\ }}\href {\doibase 10.1126/science.1188172} {\bibfield
  {journal} {\bibinfo  {journal} {Science}\ }\textbf {\bibinfo {volume}
  {328}},\ \bibinfo {pages} {879} (\bibinfo {year} {2010})}\BibitemShut
  {NoStop}%
\bibitem [{\citenamefont {Gr\"un}\ \emph
  {et~al.}(2022{\natexlab{a}})\citenamefont {Gr\"un}, \citenamefont {Ymai},
  \citenamefont {Wittmann~W.}, \citenamefont {Tonel}, \citenamefont
  {Foerster},\ and\ \citenamefont {Links}}]{atominter}%
  \BibitemOpen
  \bibfield  {author} {\bibinfo {author} {\bibfnamefont {D.~S.}\ \bibnamefont
  {Gr\"un}}, \bibinfo {author} {\bibfnamefont {L.~H.}\ \bibnamefont {Ymai}},
  \bibinfo {author} {\bibfnamefont {K.}~\bibnamefont {Wittmann~W.}}, \bibinfo
  {author} {\bibfnamefont {A.~P.}\ \bibnamefont {Tonel}}, \bibinfo {author}
  {\bibfnamefont {A.}~\bibnamefont {Foerster}}, \ and\ \bibinfo {author}
  {\bibfnamefont {J.}~\bibnamefont {Links}},\ }\bibfield  {title} {\emph
  {\bibinfo {title} {Integrable atomtronic interferometry},\ }}\href {\doibase
  10.1103/PhysRevLett.129.020401} {\bibfield  {journal} {\bibinfo  {journal}
  {Phys. Rev. Lett.}\ }\textbf {\bibinfo {volume} {129}},\ \bibinfo {pages}
  {020401} (\bibinfo {year} {2022}{\natexlab{a}})}\BibitemShut {NoStop}%
\bibitem [{\citenamefont {Gr\"un}\ \emph
  {et~al.}(2022{\natexlab{b}})\citenamefont {Gr\"un}, \citenamefont
  {Wittmann~W.}, \citenamefont {Ymai}, \citenamefont {Links},\ and\
  \citenamefont {Foerster}}]{designnoon}%
  \BibitemOpen
  \bibfield  {author} {\bibinfo {author} {\bibfnamefont {D.~S.}\ \bibnamefont
  {Gr\"un}}, \bibinfo {author} {\bibfnamefont {K.}~\bibnamefont {Wittmann~W.}},
  \bibinfo {author} {\bibfnamefont {L.~H.}\ \bibnamefont {Ymai}}, \bibinfo
  {author} {\bibfnamefont {J.}~\bibnamefont {Links}}, \ and\ \bibinfo {author}
  {\bibfnamefont {A.}~\bibnamefont {Foerster}},\ }\bibfield  {title} {\emph
  {\bibinfo {title} {Protocol designs for noon states},\ }}\href {\doibase
  10.1038/s42005-022-00812-7} {\bibfield  {journal} {\bibinfo  {journal}
  {Commun. Phys.}\ }\textbf {\bibinfo {volume} {5}},\ \bibinfo {pages} {36}
  (\bibinfo {year} {2022}{\natexlab{b}})}\BibitemShut {NoStop}%
\bibitem [{\citenamefont {Zhang}\ \emph {et~al.}(2018)\citenamefont {Zhang},
  \citenamefont {Um}, \citenamefont {Lv}, \citenamefont {Zhang}, \citenamefont
  {Duan},\ and\ \citenamefont {Kim}}]{ionnoon}%
  \BibitemOpen
  \bibfield  {author} {\bibinfo {author} {\bibfnamefont {J.}~\bibnamefont
  {Zhang}}, \bibinfo {author} {\bibfnamefont {M.}~\bibnamefont {Um}}, \bibinfo
  {author} {\bibfnamefont {D.}~\bibnamefont {Lv}}, \bibinfo {author}
  {\bibfnamefont {J.-N.}\ \bibnamefont {Zhang}}, \bibinfo {author}
  {\bibfnamefont {L.-M.}\ \bibnamefont {Duan}}, \ and\ \bibinfo {author}
  {\bibfnamefont {K.}~\bibnamefont {Kim}},\ }\bibfield  {title} {\emph
  {\bibinfo {title} {Noon states of nine quantized vibrations in two radial
  modes of a trapped ion},\ }}\href {\doibase 10.1103/PhysRevLett.121.160502}
  {\bibfield  {journal} {\bibinfo  {journal} {Phys. Rev. Lett.}\ }\textbf
  {\bibinfo {volume} {121}},\ \bibinfo {pages} {160502} (\bibinfo {year}
  {2018})}\BibitemShut {NoStop}%
\bibitem [{\citenamefont {Chen}\ and\ \citenamefont {Wei}(2017)}]{noon10}%
  \BibitemOpen
  \bibfield  {author} {\bibinfo {author} {\bibfnamefont {J.}~\bibnamefont
  {Chen}}\ and\ \bibinfo {author} {\bibfnamefont {L.~F.}\ \bibnamefont {Wei}},\
  }\bibfield  {title} {\emph {\bibinfo {title} {Deterministic generations of
  photonic noon states in cavities via shortcuts to adiabaticity},\ }}\href
  {\doibase 10.1103/PhysRevA.95.033838} {\bibfield  {journal} {\bibinfo
  {journal} {Phys. Rev. A}\ }\textbf {\bibinfo {volume} {95}},\ \bibinfo
  {pages} {033838} (\bibinfo {year} {2017})}\BibitemShut {NoStop}%
\bibitem [{\citenamefont {Lachance-Quirion}\ \emph {et~al.}(2019)\citenamefont
  {Lachance-Quirion}, \citenamefont {Tabuchi}, \citenamefont {Gloppe},
  \citenamefont {Usami},\ and\ \citenamefont {Nakamura}}]{magnon}%
  \BibitemOpen
  \bibfield  {author} {\bibinfo {author} {\bibfnamefont {D.}~\bibnamefont
  {Lachance-Quirion}}, \bibinfo {author} {\bibfnamefont {Y.}~\bibnamefont
  {Tabuchi}}, \bibinfo {author} {\bibfnamefont {A.}~\bibnamefont {Gloppe}},
  \bibinfo {author} {\bibfnamefont {K.}~\bibnamefont {Usami}}, \ and\ \bibinfo
  {author} {\bibfnamefont {Y.}~\bibnamefont {Nakamura}},\ }\bibfield  {title}
  {\emph {\bibinfo {title} {Hybrid quantum systems based on magnonics},\
  }}\href {\doibase 10.7567/1882-0786/ab248d} {\bibfield  {journal} {\bibinfo
  {journal} {Appl. Phys. Express}\ }\textbf {\bibinfo {volume} {12}},\ \bibinfo
  {pages} {070101} (\bibinfo {year} {2019})}\BibitemShut {NoStop}%
\bibitem [{\citenamefont {Li}\ \emph {et~al.}(2020)\citenamefont {Li},
  \citenamefont {Zhang}, \citenamefont {Tyberkevych}, \citenamefont {Kwok},\
  and\ \citenamefont {Novosad}}]{magnon2}%
  \BibitemOpen
  \bibfield  {author} {\bibinfo {author} {\bibfnamefont {Y.}~\bibnamefont
  {Li}}, \bibinfo {author} {\bibfnamefont {W.}~\bibnamefont {Zhang}}, \bibinfo
  {author} {\bibfnamefont {V.}~\bibnamefont {Tyberkevych}}, \bibinfo {author}
  {\bibfnamefont {W.~K.}\ \bibnamefont {Kwok}}, \ and\ \bibinfo {author}
  {\bibfnamefont {V.}~\bibnamefont {Novosad}},\ }\bibfield  {title} {\emph
  {\bibinfo {title} {Hybrid magnonics: physics, circuits, and applications for
  coherent information processing},\ }}\href {\doibase 10.1063/5.0020277}
  {\bibfield  {journal} {\bibinfo  {journal} {J. Appl. Phys.}\ }\textbf
  {\bibinfo {volume} {128}},\ \bibinfo {pages} {130902} (\bibinfo {year}
  {2020})}\BibitemShut {NoStop}%
\bibitem [{\citenamefont {Yuan}\ \emph {et~al.}(2022)\citenamefont {Yuan},
  \citenamefont {Cao}, \citenamefont {Kamra}, \citenamefont {Duine},\ and\
  \citenamefont {Yan}}]{magnon3}%
  \BibitemOpen
  \bibfield  {author} {\bibinfo {author} {\bibfnamefont {H.}~\bibnamefont
  {Yuan}}, \bibinfo {author} {\bibfnamefont {Y.}~\bibnamefont {Cao}}, \bibinfo
  {author} {\bibfnamefont {A.}~\bibnamefont {Kamra}}, \bibinfo {author}
  {\bibfnamefont {R.~A.}\ \bibnamefont {Duine}}, \ and\ \bibinfo {author}
  {\bibfnamefont {P.}~\bibnamefont {Yan}},\ }\bibfield  {title} {\emph
  {\bibinfo {title} {Quantum magnonics: When magnon spintronics meets quantum
  information science},\ }}\href {\doibase
  https://doi.org/10.1016/j.physrep.2022.03.002} {\bibfield  {journal}
  {\bibinfo  {journal} {Phys. Rep.}\ }\textbf {\bibinfo {volume} {965}},\
  \bibinfo {pages} {1} (\bibinfo {year} {2022})}\BibitemShut {NoStop}%
\bibitem [{\citenamefont {Soykal}\ and\ \citenamefont
  {Flatt\'e}(2010{\natexlab{a}})}]{nanocavity}%
  \BibitemOpen
  \bibfield  {author} {\bibinfo {author} {\bibfnamefont {O.~O.}\ \bibnamefont
  {Soykal}}\ and\ \bibinfo {author} {\bibfnamefont {M.~E.}\ \bibnamefont
  {Flatt\'e}},\ }\bibfield  {title} {\emph {\bibinfo {title} {Strong field
  interactions between a nanomagnet and a photonic cavity},\ }}\href {\doibase
  10.1103/PhysRevLett.104.077202} {\bibfield  {journal} {\bibinfo  {journal}
  {Phys. Rev. Lett.}\ }\textbf {\bibinfo {volume} {104}},\ \bibinfo {pages}
  {077202} (\bibinfo {year} {2010}{\natexlab{a}})}\BibitemShut {NoStop}%
\bibitem [{\citenamefont {Soykal}\ and\ \citenamefont
  {Flatt\'e}(2010{\natexlab{b}})}]{size}%
  \BibitemOpen
  \bibfield  {author} {\bibinfo {author} {\bibfnamefont {O.~O.}\ \bibnamefont
  {Soykal}}\ and\ \bibinfo {author} {\bibfnamefont {M.~E.}\ \bibnamefont
  {Flatt\'e}},\ }\bibfield  {title} {\emph {\bibinfo {title} {Size dependence
  of strong coupling between nanomagnets and photonic cavities},\ }}\href
  {\doibase 10.1103/PhysRevB.82.104413} {\bibfield  {journal} {\bibinfo
  {journal} {Phys. Rev. B}\ }\textbf {\bibinfo {volume} {82}},\ \bibinfo
  {pages} {104413} (\bibinfo {year} {2010}{\natexlab{b}})}\BibitemShut
  {NoStop}%
\bibitem [{\citenamefont {Li}\ \emph {et~al.}(2018)\citenamefont {Li},
  \citenamefont {Zhu},\ and\ \citenamefont {Agarwal}}]{mppentangle}%
  \BibitemOpen
  \bibfield  {author} {\bibinfo {author} {\bibfnamefont {J.}~\bibnamefont
  {Li}}, \bibinfo {author} {\bibfnamefont {S.-Y.}\ \bibnamefont {Zhu}}, \ and\
  \bibinfo {author} {\bibfnamefont {G.~S.}\ \bibnamefont {Agarwal}},\
  }\bibfield  {title} {\emph {\bibinfo {title} {Magnon-photon-phonon
  entanglement in cavity magnomechanics},\ }}\href {\doibase
  10.1103/PhysRevLett.121.203601} {\bibfield  {journal} {\bibinfo  {journal}
  {Phys. Rev. Lett.}\ }\textbf {\bibinfo {volume} {121}},\ \bibinfo {pages}
  {203601} (\bibinfo {year} {2018})}\BibitemShut {NoStop}%
\bibitem [{\citenamefont {Wang}\ \emph {et~al.}(2018)\citenamefont {Wang},
  \citenamefont {Zhang}, \citenamefont {Zhang}, \citenamefont {Li},
  \citenamefont {Hu},\ and\ \citenamefont {You}}]{bistability}%
  \BibitemOpen
  \bibfield  {author} {\bibinfo {author} {\bibfnamefont {Y.-P.}\ \bibnamefont
  {Wang}}, \bibinfo {author} {\bibfnamefont {G.-Q.}\ \bibnamefont {Zhang}},
  \bibinfo {author} {\bibfnamefont {D.}~\bibnamefont {Zhang}}, \bibinfo
  {author} {\bibfnamefont {T.-F.}\ \bibnamefont {Li}}, \bibinfo {author}
  {\bibfnamefont {C.-M.}\ \bibnamefont {Hu}}, \ and\ \bibinfo {author}
  {\bibfnamefont {J.~Q.}\ \bibnamefont {You}},\ }\bibfield  {title} {\emph
  {\bibinfo {title} {Bistability of cavity magnon polaritons},\ }}\href
  {\doibase 10.1103/PhysRevLett.120.057202} {\bibfield  {journal} {\bibinfo
  {journal} {Phys. Rev. Lett.}\ }\textbf {\bibinfo {volume} {120}},\ \bibinfo
  {pages} {057202} (\bibinfo {year} {2018})}\BibitemShut {NoStop}%
\bibitem [{\citenamefont {Zhang}\ \emph {et~al.}(2016)\citenamefont {Zhang},
  \citenamefont {Zou}, \citenamefont {Jiang},\ and\ \citenamefont
  {Tang}}]{cavitymagnonmechan}%
  \BibitemOpen
  \bibfield  {author} {\bibinfo {author} {\bibfnamefont {X.}~\bibnamefont
  {Zhang}}, \bibinfo {author} {\bibfnamefont {C.-L.}\ \bibnamefont {Zou}},
  \bibinfo {author} {\bibfnamefont {L.}~\bibnamefont {Jiang}}, \ and\ \bibinfo
  {author} {\bibfnamefont {H.}~\bibnamefont {Tang}},\ }\bibfield  {title}
  {\emph {\bibinfo {title} {Cavity magnonmechanics},\ }}\href
  {https://www.science.org/doi/10.1126/sciadv.1501286} {\bibfield  {journal}
  {\bibinfo  {journal} {Sci. Adv.}\ }\textbf {\bibinfo {volume} {2}},\ \bibinfo
  {pages} {e1501286} (\bibinfo {year} {2016})}\BibitemShut {NoStop}%
\bibitem [{\citenamefont {Tabuchi}\ \emph {et~al.}(2015)\citenamefont
  {Tabuchi}, \citenamefont {Ishino}, \citenamefont {Noguchi}, \citenamefont
  {Ishikawa}, \citenamefont {Yamazaki}, \citenamefont {Usami},\ and\
  \citenamefont {Nakamura}}]{magnonqubit1}%
  \BibitemOpen
  \bibfield  {author} {\bibinfo {author} {\bibfnamefont {Y.}~\bibnamefont
  {Tabuchi}}, \bibinfo {author} {\bibfnamefont {S.}~\bibnamefont {Ishino}},
  \bibinfo {author} {\bibfnamefont {A.}~\bibnamefont {Noguchi}}, \bibinfo
  {author} {\bibfnamefont {T.}~\bibnamefont {Ishikawa}}, \bibinfo {author}
  {\bibfnamefont {R.}~\bibnamefont {Yamazaki}}, \bibinfo {author}
  {\bibfnamefont {K.}~\bibnamefont {Usami}}, \ and\ \bibinfo {author}
  {\bibfnamefont {Y.}~\bibnamefont {Nakamura}},\ }\bibfield  {title} {\emph
  {\bibinfo {title} {Coherent coupling between a ferromagnetic magnon and a
  superconducting qubit},\ }}\href
  {https://www.science.org/lookup/doi/10.1126/science.aaa3693} {\bibfield
  {journal} {\bibinfo  {journal} {Science}\ }\textbf {\bibinfo {volume}
  {349}},\ \bibinfo {pages} {405} (\bibinfo {year} {2015})}\BibitemShut
  {NoStop}%
\bibitem [{\citenamefont {Lachance-Quirion}\ \emph {et~al.}(2020)\citenamefont
  {Lachance-Quirion}, \citenamefont {Piotr~Wolski}, \citenamefont {Tabuchi},
  \citenamefont {Kono}, \citenamefont {Usami},\ and\ \citenamefont
  {Nakamura}}]{magnonqubit2}%
  \BibitemOpen
  \bibfield  {author} {\bibinfo {author} {\bibfnamefont {D.}~\bibnamefont
  {Lachance-Quirion}}, \bibinfo {author} {\bibfnamefont {S.}~\bibnamefont
  {Piotr~Wolski}}, \bibinfo {author} {\bibfnamefont {Y.}~\bibnamefont
  {Tabuchi}}, \bibinfo {author} {\bibfnamefont {S.}~\bibnamefont {Kono}},
  \bibinfo {author} {\bibfnamefont {K.}~\bibnamefont {Usami}}, \ and\ \bibinfo
  {author} {\bibfnamefont {Y.}~\bibnamefont {Nakamura}},\ }\bibfield  {title}
  {\emph {\bibinfo {title} {Entanglement-based single-shot detection of a
  single magnon with a superconducting qubit},\ }}\href
  {https://www.science.org/lookup/doi/10.1126/science.aaz9236} {\bibfield
  {journal} {\bibinfo  {journal} {Science}\ }\textbf {\bibinfo {volume}
  {367}},\ \bibinfo {pages} {425} (\bibinfo {year} {2020})}\BibitemShut
  {NoStop}%
\bibitem [{\citenamefont {Tabuchi}\ \emph {et~al.}(2014)\citenamefont
  {Tabuchi}, \citenamefont {Ishino}, \citenamefont {Ishikawa}, \citenamefont
  {Yamazaki}, \citenamefont {Usami},\ and\ \citenamefont
  {Nakamura}}]{yigcavity1}%
  \BibitemOpen
  \bibfield  {author} {\bibinfo {author} {\bibfnamefont {Y.}~\bibnamefont
  {Tabuchi}}, \bibinfo {author} {\bibfnamefont {S.}~\bibnamefont {Ishino}},
  \bibinfo {author} {\bibfnamefont {T.}~\bibnamefont {Ishikawa}}, \bibinfo
  {author} {\bibfnamefont {R.}~\bibnamefont {Yamazaki}}, \bibinfo {author}
  {\bibfnamefont {K.}~\bibnamefont {Usami}}, \ and\ \bibinfo {author}
  {\bibfnamefont {Y.}~\bibnamefont {Nakamura}},\ }\bibfield  {title} {\emph
  {\bibinfo {title} {Hybridizing ferromagnetic magnons and microwave photons in
  the quantum limit},\ }}\href {\doibase 10.1103/PhysRevLett.113.083603}
  {\bibfield  {journal} {\bibinfo  {journal} {Phys. Rev. Lett.}\ }\textbf
  {\bibinfo {volume} {113}},\ \bibinfo {pages} {083603} (\bibinfo {year}
  {2014})}\BibitemShut {NoStop}%
\bibitem [{\citenamefont {Zhang}\ \emph {et~al.}(2014)\citenamefont {Zhang},
  \citenamefont {Zou}, \citenamefont {Jiang},\ and\ \citenamefont
  {Tang}}]{yigcavity2}%
  \BibitemOpen
  \bibfield  {author} {\bibinfo {author} {\bibfnamefont {X.}~\bibnamefont
  {Zhang}}, \bibinfo {author} {\bibfnamefont {C.-L.}\ \bibnamefont {Zou}},
  \bibinfo {author} {\bibfnamefont {L.}~\bibnamefont {Jiang}}, \ and\ \bibinfo
  {author} {\bibfnamefont {H.~X.}\ \bibnamefont {Tang}},\ }\bibfield  {title}
  {\emph {\bibinfo {title} {Strongly coupled magnons and cavity microwave
  photons},\ }}\href {\doibase 10.1103/PhysRevLett.113.156401} {\bibfield
  {journal} {\bibinfo  {journal} {Phys. Rev. Lett.}\ }\textbf {\bibinfo
  {volume} {113}},\ \bibinfo {pages} {156401} (\bibinfo {year}
  {2014})}\BibitemShut {NoStop}%
\bibitem [{\citenamefont {Qi}\ and\ \citenamefont
  {Jing}(2022{\natexlab{a}})}]{yigcavitybell2}%
  \BibitemOpen
  \bibfield  {author} {\bibinfo {author} {\bibfnamefont {S.-f.}\ \bibnamefont
  {Qi}}\ and\ \bibinfo {author} {\bibfnamefont {J.}~\bibnamefont {Jing}},\
  }\bibfield  {title} {\emph {\bibinfo {title} {Generation of bell and
  greenberger-horne-zeilinger states from a hybrid qubit-photon-magnon
  system},\ }}\href {\doibase 10.1103/PhysRevA.105.022624} {\bibfield
  {journal} {\bibinfo  {journal} {Phys. Rev. A}\ }\textbf {\bibinfo {volume}
  {105}},\ \bibinfo {pages} {022624} (\bibinfo {year}
  {2022}{\natexlab{a}})}\BibitemShut {NoStop}%
\bibitem [{\citenamefont {Yuan}\ \emph {et~al.}(2020)\citenamefont {Yuan},
  \citenamefont {Yan}, \citenamefont {Zheng}, \citenamefont {He}, \citenamefont
  {Xia},\ and\ \citenamefont {Yung}}]{yigcavitybell}%
  \BibitemOpen
  \bibfield  {author} {\bibinfo {author} {\bibfnamefont {H.~Y.}\ \bibnamefont
  {Yuan}}, \bibinfo {author} {\bibfnamefont {P.}~\bibnamefont {Yan}}, \bibinfo
  {author} {\bibfnamefont {S.}~\bibnamefont {Zheng}}, \bibinfo {author}
  {\bibfnamefont {Q.~Y.}\ \bibnamefont {He}}, \bibinfo {author} {\bibfnamefont
  {K.}~\bibnamefont {Xia}}, \ and\ \bibinfo {author} {\bibfnamefont {M.-H.}\
  \bibnamefont {Yung}},\ }\bibfield  {title} {\emph {\bibinfo {title} {Steady
  bell state generation via magnon-photon coupling},\ }}\href {\doibase
  10.1103/PhysRevLett.124.053602} {\bibfield  {journal} {\bibinfo  {journal}
  {Phys. Rev. Lett.}\ }\textbf {\bibinfo {volume} {124}},\ \bibinfo {pages}
  {053602} (\bibinfo {year} {2020})}\BibitemShut {NoStop}%
\bibitem [{\citenamefont {Blais}\ \emph {et~al.}(2021)\citenamefont {Blais},
  \citenamefont {Grimsmo}, \citenamefont {Girvin},\ and\ \citenamefont
  {Wallraff}}]{circuit}%
  \BibitemOpen
  \bibfield  {author} {\bibinfo {author} {\bibfnamefont {A.}~\bibnamefont
  {Blais}}, \bibinfo {author} {\bibfnamefont {A.~L.}\ \bibnamefont {Grimsmo}},
  \bibinfo {author} {\bibfnamefont {S.~M.}\ \bibnamefont {Girvin}}, \ and\
  \bibinfo {author} {\bibfnamefont {A.}~\bibnamefont {Wallraff}},\ }\bibfield
  {title} {\emph {\bibinfo {title} {Circuit quantum electrodynamics},\ }}\href
  {\doibase 10.1103/RevModPhys.93.025005} {\bibfield  {journal} {\bibinfo
  {journal} {Rev. Mod. Phys.}\ }\textbf {\bibinfo {volume} {93}},\ \bibinfo
  {pages} {025005} (\bibinfo {year} {2021})}\BibitemShut {NoStop}%
\bibitem [{\citenamefont {Kounalakis}\ \emph {et~al.}(2022)\citenamefont
  {Kounalakis}, \citenamefont {Bauer},\ and\ \citenamefont
  {Blanter}}]{magnonqubit}%
  \BibitemOpen
  \bibfield  {author} {\bibinfo {author} {\bibfnamefont {M.}~\bibnamefont
  {Kounalakis}}, \bibinfo {author} {\bibfnamefont {G.~E.~W.}\ \bibnamefont
  {Bauer}}, \ and\ \bibinfo {author} {\bibfnamefont {Y.~M.}\ \bibnamefont
  {Blanter}},\ }\bibfield  {title} {\emph {\bibinfo {title} {Analog quantum
  control of magnonic cat states on a chip by a superconducting qubit},\
  }}\href {\doibase 10.1103/PhysRevLett.129.037205} {\bibfield  {journal}
  {\bibinfo  {journal} {Phys. Rev. Lett.}\ }\textbf {\bibinfo {volume} {129}},\
  \bibinfo {pages} {037205} (\bibinfo {year} {2022})}\BibitemShut {NoStop}%
\bibitem [{\citenamefont {Lodahl}\ \emph {et~al.}(2017)\citenamefont {Lodahl},
  \citenamefont {Mahmoodian}, \citenamefont {Stobbe}, \citenamefont
  {Rauschenbeutel}, \citenamefont {Schneeweiss}, \citenamefont {Volz},
  \citenamefont {Pichler},\ and\ \citenamefont {Zoller}}]{chiral}%
  \BibitemOpen
  \bibfield  {author} {\bibinfo {author} {\bibfnamefont {P.}~\bibnamefont
  {Lodahl}}, \bibinfo {author} {\bibfnamefont {S.}~\bibnamefont {Mahmoodian}},
  \bibinfo {author} {\bibfnamefont {S.}~\bibnamefont {Stobbe}}, \bibinfo
  {author} {\bibfnamefont {A.}~\bibnamefont {Rauschenbeutel}}, \bibinfo
  {author} {\bibfnamefont {P.}~\bibnamefont {Schneeweiss}}, \bibinfo {author}
  {\bibfnamefont {J.}~\bibnamefont {Volz}}, \bibinfo {author} {\bibfnamefont
  {H.}~\bibnamefont {Pichler}}, \ and\ \bibinfo {author} {\bibfnamefont
  {P.}~\bibnamefont {Zoller}},\ }\bibfield  {title} {\emph {\bibinfo {title}
  {Chiral quantum optics},\ }}\href {\doibase 10.1038/nature21037} {\bibfield
  {journal} {\bibinfo  {journal} {Nature}\ }\textbf {\bibinfo {volume} {541}},\
  \bibinfo {pages} {473} (\bibinfo {year} {2017})}\BibitemShut {NoStop}%
\bibitem [{\citenamefont {Zhu}\ \emph {et~al.}(2018)\citenamefont {Zhu},
  \citenamefont {Yi}, \citenamefont {Li}, \citenamefont {Xiao}, \citenamefont
  {Zhang}, \citenamefont {Yang}, \citenamefont {Kaindl}, \citenamefont {Li},
  \citenamefont {Wang},\ and\ \citenamefont {Zhang}}]{chiral2}%
  \BibitemOpen
  \bibfield  {author} {\bibinfo {author} {\bibfnamefont {H.}~\bibnamefont
  {Zhu}}, \bibinfo {author} {\bibfnamefont {J.}~\bibnamefont {Yi}}, \bibinfo
  {author} {\bibfnamefont {M.-Y.}\ \bibnamefont {Li}}, \bibinfo {author}
  {\bibfnamefont {J.}~\bibnamefont {Xiao}}, \bibinfo {author} {\bibfnamefont
  {L.}~\bibnamefont {Zhang}}, \bibinfo {author} {\bibfnamefont {C.-W.}\
  \bibnamefont {Yang}}, \bibinfo {author} {\bibfnamefont {R.~A.}\ \bibnamefont
  {Kaindl}}, \bibinfo {author} {\bibfnamefont {L.-J.}\ \bibnamefont {Li}},
  \bibinfo {author} {\bibfnamefont {Y.}~\bibnamefont {Wang}}, \ and\ \bibinfo
  {author} {\bibfnamefont {X.}~\bibnamefont {Zhang}},\ }\bibfield  {title}
  {\emph {\bibinfo {title} {Observation of chiral phonons},\ }}\href {\doibase
  10.1126/science.aar2711} {\bibfield  {journal} {\bibinfo  {journal}
  {Science}\ }\textbf {\bibinfo {volume} {359}},\ \bibinfo {pages} {579}
  (\bibinfo {year} {2018})}\BibitemShut {NoStop}%
\bibitem [{\citenamefont {S\o{}rensen}\ \emph {et~al.}(2005)\citenamefont
  {S\o{}rensen}, \citenamefont {Demler},\ and\ \citenamefont
  {Lukin}}]{fraction}%
  \BibitemOpen
  \bibfield  {author} {\bibinfo {author} {\bibfnamefont {A.~S.}\ \bibnamefont
  {S\o{}rensen}}, \bibinfo {author} {\bibfnamefont {E.}~\bibnamefont {Demler}},
  \ and\ \bibinfo {author} {\bibfnamefont {M.~D.}\ \bibnamefont {Lukin}},\
  }\bibfield  {title} {\emph {\bibinfo {title} {Fractional quantum hall states
  of atoms in optical lattices},\ }}\href {\doibase
  10.1103/PhysRevLett.94.086803} {\bibfield  {journal} {\bibinfo  {journal}
  {Phys. Rev. Lett.}\ }\textbf {\bibinfo {volume} {94}},\ \bibinfo {pages}
  {086803} (\bibinfo {year} {2005})}\BibitemShut {NoStop}%
\bibitem [{\citenamefont {Koch}\ \emph {et~al.}(2010)\citenamefont {Koch},
  \citenamefont {Houck}, \citenamefont {Hur},\ and\ \citenamefont
  {Girvin}}]{timereversal}%
  \BibitemOpen
  \bibfield  {author} {\bibinfo {author} {\bibfnamefont {J.}~\bibnamefont
  {Koch}}, \bibinfo {author} {\bibfnamefont {A.~A.}\ \bibnamefont {Houck}},
  \bibinfo {author} {\bibfnamefont {K.~L.}\ \bibnamefont {Hur}}, \ and\
  \bibinfo {author} {\bibfnamefont {S.~M.}\ \bibnamefont {Girvin}},\ }\bibfield
   {title} {\emph {\bibinfo {title} {Time-reversal-symmetry breaking in
  circuit-qed-based photon lattices},\ }}\href {\doibase
  10.1103/PhysRevA.82.043811} {\bibfield  {journal} {\bibinfo  {journal} {Phys.
  Rev. A}\ }\textbf {\bibinfo {volume} {82}},\ \bibinfo {pages} {043811}
  (\bibinfo {year} {2010})}\BibitemShut {NoStop}%
\bibitem [{\citenamefont {Roushan1}\ \emph {et~al.}(2017)\citenamefont
  {Roushan1}, \citenamefont {Neill}, \citenamefont {Megrant}, \citenamefont
  {Chen}, \citenamefont {Babbush}, \citenamefont {Barends}, \citenamefont
  {Campbell}, \citenamefont {Chen}, \citenamefont {Chiaro}, \citenamefont
  {Dunsworth}, \citenamefont {Fowler}, \citenamefont {Jeffrey}, \citenamefont
  {Kelly}, \citenamefont {Lucero}, \citenamefont {Mutus}, \citenamefont
  {\'OMalley}, \citenamefont {Neeley}, \citenamefont {Quintana}, \citenamefont
  {Sank}, \citenamefont {Vainsencher}, \citenamefont {Wenner}, \citenamefont
  {White}, \citenamefont {Kapit}, \citenamefont {Neven},\ and\ \citenamefont
  {Martinis}}]{chiralcurrent}%
  \BibitemOpen
  \bibfield  {author} {\bibinfo {author} {\bibfnamefont {P.}~\bibnamefont
  {Roushan1}}, \bibinfo {author} {\bibfnamefont {C.}~\bibnamefont {Neill}},
  \bibinfo {author} {\bibfnamefont {A.}~\bibnamefont {Megrant}}, \bibinfo
  {author} {\bibfnamefont {Y.}~\bibnamefont {Chen}}, \bibinfo {author}
  {\bibfnamefont {R.}~\bibnamefont {Babbush}}, \bibinfo {author} {\bibfnamefont
  {R.}~\bibnamefont {Barends}}, \bibinfo {author} {\bibfnamefont
  {B.}~\bibnamefont {Campbell}}, \bibinfo {author} {\bibfnamefont
  {Z.}~\bibnamefont {Chen}}, \bibinfo {author} {\bibfnamefont {B.}~\bibnamefont
  {Chiaro}}, \bibinfo {author} {\bibfnamefont {A.}~\bibnamefont {Dunsworth}},
  \bibinfo {author} {\bibfnamefont {A.}~\bibnamefont {Fowler}}, \bibinfo
  {author} {\bibfnamefont {E.}~\bibnamefont {Jeffrey}}, \bibinfo {author}
  {\bibfnamefont {J.}~\bibnamefont {Kelly}}, \bibinfo {author} {\bibfnamefont
  {E.}~\bibnamefont {Lucero}}, \bibinfo {author} {\bibfnamefont
  {J.}~\bibnamefont {Mutus}}, \bibinfo {author} {\bibfnamefont
  {P.}~\bibnamefont {\'OMalley}}, \bibinfo {author} {\bibfnamefont
  {M.}~\bibnamefont {Neeley}}, \bibinfo {author} {\bibfnamefont
  {C.}~\bibnamefont {Quintana}}, \bibinfo {author} {\bibfnamefont
  {D.}~\bibnamefont {Sank}}, \bibinfo {author} {\bibfnamefont {A.}~\bibnamefont
  {Vainsencher}}, \bibinfo {author} {\bibfnamefont {J.}~\bibnamefont {Wenner}},
  \bibinfo {author} {\bibfnamefont {T.}~\bibnamefont {White}}, \bibinfo
  {author} {\bibfnamefont {E.}~\bibnamefont {Kapit}}, \bibinfo {author}
  {\bibfnamefont {H.}~\bibnamefont {Neven}}, \ and\ \bibinfo {author}
  {\bibfnamefont {J.}~\bibnamefont {Martinis}},\ }\bibfield  {title} {\emph
  {\bibinfo {title} {Chiral ground-state currents of interacting photons in a
  synthetic magnetic field},\ }}\href {\doibase 10.1038/NPHYS3930} {\bibfield
  {journal} {\bibinfo  {journal} {Nat. Phys.}\ }\textbf {\bibinfo {volume}
  {13}},\ \bibinfo {pages} {146} (\bibinfo {year} {2017})}\BibitemShut
  {NoStop}%
\bibitem [{\citenamefont {Wang}\ \emph {et~al.}(2016)\citenamefont {Wang},
  \citenamefont {Cai}, \citenamefont {Liu},\ and\ \citenamefont
  {Scully}}]{floquetnoon}%
  \BibitemOpen
  \bibfield  {author} {\bibinfo {author} {\bibfnamefont {D.-W.}\ \bibnamefont
  {Wang}}, \bibinfo {author} {\bibfnamefont {H.}~\bibnamefont {Cai}}, \bibinfo
  {author} {\bibfnamefont {R.-B.}\ \bibnamefont {Liu}}, \ and\ \bibinfo
  {author} {\bibfnamefont {M.~O.}\ \bibnamefont {Scully}},\ }\bibfield  {title}
  {\emph {\bibinfo {title} {Mesoscopic superposition states generated by
  synthetic spin-orbit interaction in {F}ock-state lattices},\ }}\href
  {\doibase 10.1103/PhysRevLett.116.220502} {\bibfield  {journal} {\bibinfo
  {journal} {Phys. Rev. Lett.}\ }\textbf {\bibinfo {volume} {116}},\ \bibinfo
  {pages} {220502} (\bibinfo {year} {2016})}\BibitemShut {NoStop}%
\bibitem [{\citenamefont {Wang}\ \emph {et~al.}(2019)\citenamefont {Wang},
  \citenamefont {Song}, \citenamefont {Feng}, \citenamefont {Cai},
  \citenamefont {Xu}, \citenamefont {Deng}, \citenamefont {Li}, \citenamefont
  {Zheng}, \citenamefont {Zhu}, \citenamefont {Wang}, \citenamefont {Zhu},\
  and\ \citenamefont {Scully}}]{chiralspin}%
  \BibitemOpen
  \bibfield  {author} {\bibinfo {author} {\bibfnamefont {D.-W.}\ \bibnamefont
  {Wang}}, \bibinfo {author} {\bibfnamefont {C.}~\bibnamefont {Song}}, \bibinfo
  {author} {\bibfnamefont {W.}~\bibnamefont {Feng}}, \bibinfo {author}
  {\bibfnamefont {H.}~\bibnamefont {Cai}}, \bibinfo {author} {\bibfnamefont
  {D.}~\bibnamefont {Xu}}, \bibinfo {author} {\bibfnamefont {H.}~\bibnamefont
  {Deng}}, \bibinfo {author} {\bibfnamefont {H.}~\bibnamefont {Li}}, \bibinfo
  {author} {\bibfnamefont {D.}~\bibnamefont {Zheng}}, \bibinfo {author}
  {\bibfnamefont {X.}~\bibnamefont {Zhu}}, \bibinfo {author} {\bibfnamefont
  {H.}~\bibnamefont {Wang}}, \bibinfo {author} {\bibfnamefont {S.}~\bibnamefont
  {Zhu}}, \ and\ \bibinfo {author} {\bibfnamefont {M.~O.}\ \bibnamefont
  {Scully}},\ }\bibfield  {title} {\emph {\bibinfo {title} {Synthesis of
  antisymmetric spin exchange interaction and chiral spin clusters in
  superconducting circuits},\ }}\href {\doibase 10.1038/s41567-018-0400-9}
  {\bibfield  {journal} {\bibinfo  {journal} {Nat. Phys.}\ }\textbf {\bibinfo
  {volume} {15}},\ \bibinfo {pages} {382} (\bibinfo {year} {2019})}\BibitemShut
  {NoStop}%
\bibitem [{\citenamefont {Liu}\ \emph {et~al.}(2020)\citenamefont {Liu},
  \citenamefont {Feng}, \citenamefont {Ren}, \citenamefont {Wang},\ and\
  \citenamefont {Wang}}]{chiralspin2}%
  \BibitemOpen
  \bibfield  {author} {\bibinfo {author} {\bibfnamefont {W.}~\bibnamefont
  {Liu}}, \bibinfo {author} {\bibfnamefont {W.}~\bibnamefont {Feng}}, \bibinfo
  {author} {\bibfnamefont {W.}~\bibnamefont {Ren}}, \bibinfo {author}
  {\bibfnamefont {D.-W.}\ \bibnamefont {Wang}}, \ and\ \bibinfo {author}
  {\bibfnamefont {H.}~\bibnamefont {Wang}},\ }\bibfield  {title} {\emph
  {\bibinfo {title} {Synthesizing three-body interaction of spin chirality with
  superconducting qubits},\ }}\href {\doibase 10.1063/1.5140884} {\bibfield
  {journal} {\bibinfo  {journal} {Appl. Phys. Lett.}\ }\textbf {\bibinfo
  {volume} {116}},\ \bibinfo {pages} {114001} (\bibinfo {year}
  {2020})}\BibitemShut {NoStop}%
\bibitem [{\citenamefont {Goldman}\ and\ \citenamefont
  {Dalibard}(2014)}]{Floquet1}%
  \BibitemOpen
  \bibfield  {author} {\bibinfo {author} {\bibfnamefont {N.}~\bibnamefont
  {Goldman}}\ and\ \bibinfo {author} {\bibfnamefont {J.}~\bibnamefont
  {Dalibard}},\ }\bibfield  {title} {\emph {\bibinfo {title} {Periodically
  driven quantum systems: Effective hamiltonians and engineered gauge fields},\
  }}\href {\doibase 10.1103/PhysRevX.4.031027} {\bibfield  {journal} {\bibinfo
  {journal} {Phys. Rev. X}\ }\textbf {\bibinfo {volume} {4}},\ \bibinfo {pages}
  {031027} (\bibinfo {year} {2014})}\BibitemShut {NoStop}%
\bibitem [{\citenamefont {Shao}\ \emph {et~al.}(2017)\citenamefont {Shao},
  \citenamefont {Wu},\ and\ \citenamefont {Feng}}]{James}%
  \BibitemOpen
  \bibfield  {author} {\bibinfo {author} {\bibfnamefont {W.}~\bibnamefont
  {Shao}}, \bibinfo {author} {\bibfnamefont {C.}~\bibnamefont {Wu}}, \ and\
  \bibinfo {author} {\bibfnamefont {X.-L.}\ \bibnamefont {Feng}},\ }\bibfield
  {title} {\emph {\bibinfo {title} {Generalized james' effective hamiltonian
  method},\ }}\href {\doibase 10.1103/PhysRevA.95.032124} {\bibfield  {journal}
  {\bibinfo  {journal} {Phys. Rev. A}\ }\textbf {\bibinfo {volume} {95}},\
  \bibinfo {pages} {032124} (\bibinfo {year} {2017})}\BibitemShut {NoStop}%
\bibitem [{\citenamefont {Bukov}\ \emph {et~al.}(2015)\citenamefont {Bukov},
  \citenamefont {D'Alessio},\ and\ \citenamefont
  {Polkovnikov}}]{Floquettheory}%
  \BibitemOpen
  \bibfield  {author} {\bibinfo {author} {\bibfnamefont {M.}~\bibnamefont
  {Bukov}}, \bibinfo {author} {\bibfnamefont {L.}~\bibnamefont {D'Alessio}}, \
  and\ \bibinfo {author} {\bibfnamefont {A.}~\bibnamefont {Polkovnikov}},\
  }\bibfield  {title} {\emph {\bibinfo {title} {Universal high-frequency
  behavior of periodically driven system: from dynamical stabilization to
  floquet engineering},\ }}\href {\doibase 10.1080/00018732.2015.1055918}
  {\bibfield  {journal} {\bibinfo  {journal} {Adv. in Phys.}\ }\textbf
  {\bibinfo {volume} {64}},\ \bibinfo {pages} {139} (\bibinfo {year}
  {2015})}\BibitemShut {NoStop}%
\bibitem [{\citenamefont {Petiziol}\ \emph {et~al.}(2021)\citenamefont
  {Petiziol}, \citenamefont {Sameti}, \citenamefont {Carretta}, \citenamefont
  {Wimberger},\ and\ \citenamefont {Mintert}}]{Floquet2}%
  \BibitemOpen
  \bibfield  {author} {\bibinfo {author} {\bibfnamefont {F.}~\bibnamefont
  {Petiziol}}, \bibinfo {author} {\bibfnamefont {M.}~\bibnamefont {Sameti}},
  \bibinfo {author} {\bibfnamefont {S.}~\bibnamefont {Carretta}}, \bibinfo
  {author} {\bibfnamefont {S.}~\bibnamefont {Wimberger}}, \ and\ \bibinfo
  {author} {\bibfnamefont {F.}~\bibnamefont {Mintert}},\ }\bibfield  {title}
  {\emph {\bibinfo {title} {Quantum simulation of three-body interactions in
  weakly driven quantum systems},\ }}\href {\doibase
  10.1103/PhysRevLett.126.250504} {\bibfield  {journal} {\bibinfo  {journal}
  {Phys. Rev. Lett.}\ }\textbf {\bibinfo {volume} {126}},\ \bibinfo {pages}
  {250504} (\bibinfo {year} {2021})}\BibitemShut {NoStop}%
\bibitem [{\citenamefont {Wu}\ \emph {et~al.}(2018)\citenamefont {Wu},
  \citenamefont {Yang}, \citenamefont {Gong}, \citenamefont {Zheng},
  \citenamefont {Deng}, \citenamefont {Yan}, \citenamefont {Zhao},
  \citenamefont {Huang}, \citenamefont {Castellano}, \citenamefont {Munro},
  \citenamefont {Nemoto}, \citenamefont {Zheng}, \citenamefont {Sun},
  \citenamefont {Liu}, \citenamefont {Zhu},\ and\ \citenamefont {Lu}}]{switch}%
  \BibitemOpen
  \bibfield  {author} {\bibinfo {author} {\bibfnamefont {Y.}~\bibnamefont
  {Wu}}, \bibinfo {author} {\bibfnamefont {L.-P.}\ \bibnamefont {Yang}},
  \bibinfo {author} {\bibfnamefont {M.}~\bibnamefont {Gong}}, \bibinfo {author}
  {\bibfnamefont {Y.}~\bibnamefont {Zheng}}, \bibinfo {author} {\bibfnamefont
  {H.}~\bibnamefont {Deng}}, \bibinfo {author} {\bibfnamefont {Z.}~\bibnamefont
  {Yan}}, \bibinfo {author} {\bibfnamefont {Y.}~\bibnamefont {Zhao}}, \bibinfo
  {author} {\bibfnamefont {K.}~\bibnamefont {Huang}}, \bibinfo {author}
  {\bibfnamefont {A.~D.}\ \bibnamefont {Castellano}}, \bibinfo {author}
  {\bibfnamefont {W.~J.}\ \bibnamefont {Munro}}, \bibinfo {author}
  {\bibfnamefont {K.}~\bibnamefont {Nemoto}}, \bibinfo {author} {\bibfnamefont
  {D.-N.}\ \bibnamefont {Zheng}}, \bibinfo {author} {\bibfnamefont
  {C.}~\bibnamefont {Sun}}, \bibinfo {author} {\bibfnamefont {Y.-x.}\
  \bibnamefont {Liu}}, \bibinfo {author} {\bibfnamefont {X.}~\bibnamefont
  {Zhu}}, \ and\ \bibinfo {author} {\bibfnamefont {L.}~\bibnamefont {Lu}},\
  }\bibfield  {title} {\emph {\bibinfo {title} {An efficient and compact switch
  for quantum circuits},\ }}\href {\doibase 10.1038/s41534-018-0099-6}
  {\bibfield  {journal} {\bibinfo  {journal} {npj. Quantum Inf.}\ }\textbf
  {\bibinfo {volume} {4}},\ \bibinfo {pages} {50} (\bibinfo {year}
  {2018})}\BibitemShut {NoStop}%
\bibitem [{\citenamefont {Kyriienko}\ and\ \citenamefont
  {S\o{}rensen}(2018)}]{Floquetsimu}%
  \BibitemOpen
  \bibfield  {author} {\bibinfo {author} {\bibfnamefont {O.}~\bibnamefont
  {Kyriienko}}\ and\ \bibinfo {author} {\bibfnamefont {A.~S.}\ \bibnamefont
  {S\o{}rensen}},\ }\bibfield  {title} {\emph {\bibinfo {title} {Floquet
  quantum simulation with superconducting qubits},\ }}\href {\doibase
  10.1103/PhysRevApplied.9.064029} {\bibfield  {journal} {\bibinfo  {journal}
  {Phys. Rev. Applied}\ }\textbf {\bibinfo {volume} {9}},\ \bibinfo {pages}
  {064029} (\bibinfo {year} {2018})}\BibitemShut {NoStop}%
\bibitem [{\citenamefont {Liberato}\ \emph {et~al.}(2007)\citenamefont
  {Liberato}, \citenamefont {Ciuti},\ and\ \citenamefont
  {Carusotto}}]{timecoupling}%
  \BibitemOpen
  \bibfield  {author} {\bibinfo {author} {\bibfnamefont {S.~D.}\ \bibnamefont
  {Liberato}}, \bibinfo {author} {\bibfnamefont {C.}~\bibnamefont {Ciuti}}, \
  and\ \bibinfo {author} {\bibfnamefont {I.}~\bibnamefont {Carusotto}},\
  }\bibfield  {title} {\emph {\bibinfo {title} {Quantum vacuum radiation
  spectra from a semiconductor microcavity with a time-modulated vacuum rabi
  frequency},\ }}\href {\doibase 10.1103/PhysRevLett.98.103602} {\bibfield
  {journal} {\bibinfo  {journal} {Phys. Rev. Lett.}\ }\textbf {\bibinfo
  {volume} {98}},\ \bibinfo {pages} {103602} (\bibinfo {year}
  {2007})}\BibitemShut {NoStop}%
\bibitem [{\citenamefont {Xu}\ \emph {et~al.}(2020)\citenamefont {Xu},
  \citenamefont {Zhong}, \citenamefont {Han}, \citenamefont {Jin},
  \citenamefont {Jiang},\ and\ \citenamefont {Zhang}}]{floquetmagnon}%
  \BibitemOpen
  \bibfield  {author} {\bibinfo {author} {\bibfnamefont {J.}~\bibnamefont
  {Xu}}, \bibinfo {author} {\bibfnamefont {C.}~\bibnamefont {Zhong}}, \bibinfo
  {author} {\bibfnamefont {X.}~\bibnamefont {Han}}, \bibinfo {author}
  {\bibfnamefont {D.}~\bibnamefont {Jin}}, \bibinfo {author} {\bibfnamefont
  {L.}~\bibnamefont {Jiang}}, \ and\ \bibinfo {author} {\bibfnamefont
  {X.}~\bibnamefont {Zhang}},\ }\bibfield  {title} {\emph {\bibinfo {title}
  {Floquet cavity electromagnonics},\ }}\href {\doibase
  10.1103/PhysRevLett.125.237201} {\bibfield  {journal} {\bibinfo  {journal}
  {Phys. Rev. Lett.}\ }\textbf {\bibinfo {volume} {125}},\ \bibinfo {pages}
  {237201} (\bibinfo {year} {2020})}\BibitemShut {NoStop}%
\bibitem [{\citenamefont {Wang}\ \emph {et~al.}(2017)\citenamefont {Wang},
  \citenamefont {Hu}, \citenamefont {Xu}, \citenamefont {Liu}, \citenamefont
  {Ma}, \citenamefont {Zheng}, \citenamefont {Vijay}, \citenamefont {Song},
  \citenamefont {Duan},\ and\ \citenamefont {Sun}}]{Fockmeasure}%
  \BibitemOpen
  \bibfield  {author} {\bibinfo {author} {\bibfnamefont {W.}~\bibnamefont
  {Wang}}, \bibinfo {author} {\bibfnamefont {L.}~\bibnamefont {Hu}}, \bibinfo
  {author} {\bibfnamefont {Y.}~\bibnamefont {Xu}}, \bibinfo {author}
  {\bibfnamefont {K.}~\bibnamefont {Liu}}, \bibinfo {author} {\bibfnamefont
  {Y.}~\bibnamefont {Ma}}, \bibinfo {author} {\bibfnamefont {S.-B.}\
  \bibnamefont {Zheng}}, \bibinfo {author} {\bibfnamefont {R.}~\bibnamefont
  {Vijay}}, \bibinfo {author} {\bibfnamefont {Y.~P.}\ \bibnamefont {Song}},
  \bibinfo {author} {\bibfnamefont {L.-M.}\ \bibnamefont {Duan}}, \ and\
  \bibinfo {author} {\bibfnamefont {L.}~\bibnamefont {Sun}},\ }\bibfield
  {title} {\emph {\bibinfo {title} {Converting quasiclassical states into
  arbitrary {F}ock state superpositions in a superconducting circuit},\ }}\href
  {\doibase 10.1103/PhysRevLett.118.223604} {\bibfield  {journal} {\bibinfo
  {journal} {Phys. Rev. Lett.}\ }\textbf {\bibinfo {volume} {118}},\ \bibinfo
  {pages} {223604} (\bibinfo {year} {2017})}\BibitemShut {NoStop}%
\bibitem [{\citenamefont {Qi}\ and\ \citenamefont
  {Jing}(2022{\natexlab{b}})}]{localphase}%
  \BibitemOpen
  \bibfield  {author} {\bibinfo {author} {\bibfnamefont {S.-f.}\ \bibnamefont
  {Qi}}\ and\ \bibinfo {author} {\bibfnamefont {J.}~\bibnamefont {Jing}},\
  }\bibfield  {title} {\emph {\bibinfo {title} {Accelerated adiabatic passage
  in cavity magnomechanics},\ }}\href {\doibase 10.1103/PhysRevA.105.053710}
  {\bibfield  {journal} {\bibinfo  {journal} {Phys. Rev. A}\ }\textbf {\bibinfo
  {volume} {105}},\ \bibinfo {pages} {053710} (\bibinfo {year}
  {2022}{\natexlab{b}})}\BibitemShut {NoStop}%
\bibitem [{\citenamefont {Ruschhaupt}\ \emph {et~al.}(2012)\citenamefont
  {Ruschhaupt}, \citenamefont {Chen}, \citenamefont {Alonso},\ and\
  \citenamefont {Muga}}]{syserrorpo}%
  \BibitemOpen
  \bibfield  {author} {\bibinfo {author} {\bibfnamefont {A.}~\bibnamefont
  {Ruschhaupt}}, \bibinfo {author} {\bibfnamefont {X.}~\bibnamefont {Chen}},
  \bibinfo {author} {\bibfnamefont {D.}~\bibnamefont {Alonso}}, \ and\ \bibinfo
  {author} {\bibfnamefont {J.~G.}\ \bibnamefont {Muga}},\ }\bibfield  {title}
  {\emph {\bibinfo {title} {Optimally robust shortcuts to population inversion
  in two-level quantum systems},\ }}\href {\doibase
  10.1088/1367-2630/14/9/093040} {\bibfield  {journal} {\bibinfo  {journal}
  {New J. Phys.}\ }\textbf {\bibinfo {volume} {14}},\ \bibinfo {pages} {093040}
  (\bibinfo {year} {2012})}\BibitemShut {NoStop}%
\bibitem [{\citenamefont {Rameshti}\ \emph {et~al.}(2022)\citenamefont
  {Rameshti}, \citenamefont {Kusminskiy}, \citenamefont {Haigh}, \citenamefont
  {Usami}, \citenamefont {Lachance-Quirion}, \citenamefont {Nakamura},
  \citenamefont {Hu}, \citenamefont {Tang}, \citenamefont {Bauer},\ and\
  \citenamefont {Blanter}}]{cavitymagnonics}%
  \BibitemOpen
  \bibfield  {author} {\bibinfo {author} {\bibfnamefont {B.~Z.}\ \bibnamefont
  {Rameshti}}, \bibinfo {author} {\bibfnamefont {S.~V.}\ \bibnamefont
  {Kusminskiy}}, \bibinfo {author} {\bibfnamefont {J.~A.}\ \bibnamefont
  {Haigh}}, \bibinfo {author} {\bibfnamefont {K.}~\bibnamefont {Usami}},
  \bibinfo {author} {\bibfnamefont {D.}~\bibnamefont {Lachance-Quirion}},
  \bibinfo {author} {\bibfnamefont {Y.}~\bibnamefont {Nakamura}}, \bibinfo
  {author} {\bibfnamefont {C.-M.}\ \bibnamefont {Hu}}, \bibinfo {author}
  {\bibfnamefont {H.~X.}\ \bibnamefont {Tang}}, \bibinfo {author}
  {\bibfnamefont {G.~E.}\ \bibnamefont {Bauer}}, \ and\ \bibinfo {author}
  {\bibfnamefont {Y.~M.}\ \bibnamefont {Blanter}},\ }\bibfield  {title} {\emph
  {\bibinfo {title} {Cavity magnonics},\ }}\href {\doibase
  10.1016/j.physrep.2022.06.001} {\bibfield  {journal} {\bibinfo  {journal}
  {Phys. Rep.}\ }\textbf {\bibinfo {volume} {979}},\ \bibinfo {pages} {1}
  (\bibinfo {year} {2022})}\BibitemShut {NoStop}%
\bibitem [{\citenamefont {Kjaergaard}\ \emph {et~al.}(2020)\citenamefont
  {Kjaergaard}, \citenamefont {Schwartz}, \citenamefont {Braum\"{u}ller},
  \citenamefont {Krantz}, \citenamefont {Wang}, \citenamefont {Gustavsson},\
  and\ \citenamefont {Oliver}}]{tunequbit}%
  \BibitemOpen
  \bibfield  {author} {\bibinfo {author} {\bibfnamefont {M.}~\bibnamefont
  {Kjaergaard}}, \bibinfo {author} {\bibfnamefont {M.~E.}\ \bibnamefont
  {Schwartz}}, \bibinfo {author} {\bibfnamefont {J.}~\bibnamefont
  {Braum\"{u}ller}}, \bibinfo {author} {\bibfnamefont {P.}~\bibnamefont
  {Krantz}}, \bibinfo {author} {\bibfnamefont {J.~I.-J.}\ \bibnamefont {Wang}},
  \bibinfo {author} {\bibfnamefont {S.}~\bibnamefont {Gustavsson}}, \ and\
  \bibinfo {author} {\bibfnamefont {W.~D.}\ \bibnamefont {Oliver}},\ }\bibfield
   {title} {\emph {\bibinfo {title} {Superconducting qubits: Current state of
  play},\ }}\href {\doibase 10.1146/annurev-conmatphys-031119-050605}
  {\bibfield  {journal} {\bibinfo  {journal} {Annu. Rev. of Conden. Matter
  Phys.}\ }\textbf {\bibinfo {volume} {11}},\ \bibinfo {pages} {369} (\bibinfo
  {year} {2020})}\BibitemShut {NoStop}%
\bibitem [{\citenamefont {Sanders}(1992)}]{entanglecoher}%
  \BibitemOpen
  \bibfield  {author} {\bibinfo {author} {\bibfnamefont {B.~C.}\ \bibnamefont
  {Sanders}},\ }\bibfield  {title} {\emph {\bibinfo {title} {Entangled coherent
  states},\ }}\href {\doibase 10.1103/PhysRevA.45.6811} {\bibfield  {journal}
  {\bibinfo  {journal} {Phys. Rev. A}\ }\textbf {\bibinfo {volume} {45}},\
  \bibinfo {pages} {6811} (\bibinfo {year} {1992})}\BibitemShut {NoStop}%
\end{thebibliography}%

\end{document}